\documentclass[%
 reprint,
 superscriptaddress,
 amsmath,amssymb,
 aps,
 prx,
]{revtex4-2}
\usepackage{physics}
\usepackage{graphicx}
\usepackage{dcolumn}
\usepackage{bm}
\usepackage{algorithmic}
\usepackage{siunitx}
\usepackage{mathtools}

\begin{document}


\title{Linear Viscoelastic Response of the Vertex Model\\ with Internal and External Dissipation: Normal Modes Analysis}
%

\author{Sijie Tong}
\affiliation{Department of Mechanical and Aerospace Engineering, Princeton University, Princeton, New Jersey 08544, USA}
\author{Rastko Sknepnek}
\email{r.sknepnek@dundee.ac.uk}
\affiliation{School of Science and Engineering, University of Dundee, Dundee DD1 4HN, United Kingdom}
\affiliation{School of Life Sciences, University of Dundee, Dundee DD1 5EH, United Kingdom}
\author{Andrej Ko{\v s}mrlj}
\email{andrej@princeton.edu}
\affiliation{Department of Mechanical and Aerospace Engineering, Princeton University, Princeton, New Jersey 08544, USA}
\affiliation{Princeton Institute of Materials, Princeton University, Princeton, New Jersey 08544, USA}


\begin{abstract}
We use the normal mode formalism to study the shear rheology of the vertex model for epithelial tissue mechanics in the overdamped linear response regime. We consider systems with external (e.g., cell-substrate) and internal (e.g., cell-cell) dissipation mechanisms, and derive expressions for stresses on cells due to mechanical and dissipative forces. The semi-analytical method developed here is, however, general and can be directly applied to study the linear response of a broad class of soft matter systems with internal and external dissipation. It involves normal mode decomposition to calculate linear loss and storage moduli of the system. Specifically, displacements along each normal mode produce stresses due to elastic deformation and internal dissipation, which are in force balance with loads due to external dissipation. Each normal mode responds with a characteristic relaxation timescale, and its rheological behavior can be described as a combination of a standard linear solid element due to elastic stresses and a Jeffreys model element due to the internal dissipative stresses. The total response of the system is then fully determined by connecting in parallel all the viscoelastic elements corresponding to individual normal modes. This allows full characterization of the potentially complex linear rheological response of the system at all driving frequencies and identification of collective excitations. We show that internal and external dissipation mechanisms lead to qualitatively different rheological behaviors due to the presence or absence of Jeffreys elements, which is particularly pronounced at high driving frequencies. Our findings, therefore, underscore the importance of microscopic dissipation mechanisms in understanding the rheological behavior of soft materials and tissues, in particular. 

\end{abstract}

\maketitle

\section{Introduction}
Soft materials exhibit rich mechanical and rheological behaviors such as viscoelasticity, time-dependent viscosity, shear thickening and thinning, etc.~\cite{larson1999structure,doi2013soft} Understanding these behaviors is of both fundamental and practical interest with many soft materials being ubiquitous in everyday life. Rich macroscopic rheological properties result from typically complex amorphous microscopic structure, relatively weak interactions between constitutive elements  (i.e., often comparable with the thermal energy), and dissipation. In polymer melts, for example, the storage modulus has a characteristic plateau over a range of frequencies due to very different time scales associated with the diffusion along and perpendicular to the polymer chain.~\cite{doi1988theory} In dense emulsions, the loss modulus remains constant under oscillatory shear for driving frequencies as low as $10^{-2}-1$ $\mathrm{Hz}$,~\cite{mason1995elasticity} in contradiction to the linear response theory which predicts that the loss modulus is an odd function of frequency, and, therefore, should vanish for at low frequencies. This behavior is attributed to slow ``glassy'' dynamics.~\cite{sollich1997rheology,sollich1998rheological} In granular media approaching the unjamming transition, the density of vibrational states remains constant at low frequencies leading to a diverging length scale,~\cite{silbert2005vibrations} which results in scaling behavior of elasticity and transport properties, and force propagation as the packing fraction approaches the critical value.~\cite{o2003jamming, xu2009energy, wyart2008elasticity} 

The mechanical and rheological properties of living matter share a lot in common with the inanimate amorphous soft active  matter.~\cite{marchetti2013hydrodynamics} Collective behavior of cells in confluent epithelial tissues and cell monolayers, for example, can be modeled as dense soft active matter and have been known to show complex collective behaviors~\cite{poujade2007collective,trepat2009physical, tambe2011collective, tlili2018collective, saraswathibhatla2021coordinated} with dynamical patterns reminiscent of supercooled liquids and active nematics.~\cite{szabo2010collective, angelini2011glass, saw2017topological, henkes2020dense} There are, however, key differences between living systems and inanimate active matter. Notably, biological systems can tune their properties in response to chemical and mechanical changes in the environment. Living systems also grow, age, regenerate, etc. The understanding of how such processes affect the physics of living matter is a very active area of research. Epithelial tissues and cell monolayers have been widely used as model systems to study complex mesoscale behavior of biological systems.~\cite{trepat2009physical, trepat2011plithotaxis, angelini2011glass, harris2012characterizing, hakim2017collective, park2015unjamming}  In particular, there has been growing interest in studying the viscoelastic response of these systems,~\cite {harris2012characterizing} which revealed complex non-linear behavior and led to interesting theoretical models, e.g., the description based on fractional derivatives to capture anomalous power-law relaxation.~\cite{bonfanti2020unified, bonfanti2020fractional}

Understanding the full non-linear response of soft and biological matter is faced with major challenges such as multiple competing length and time scales, and irreversible plastic microscopic rearrangements that produce non-local stresses, aging, etc.~\cite{nicolas2018deformation, czajkowski2018hydrodynamics, huang2022shear, grossman2022instabilities} Despite many theoretical and numerical studies of various aspects of the non-linear response, a general framework is still lacking. In the linear regime, however, many of these challenges can be neglected to make a comprehensive treatment possible. The linear response treatment is clearly inadequate if one is interested in the long-time behavior under large deformations. It can, nonetheless, provide valuable insights into often complex baseline behaviors of the system. 

In this paper, we develop a semi-analytic formalism that can be used for investigating the linear rheological response of a broad class of soft and biological matter over the full range of driving frequencies. Our approach is based on the well-known decomposition in terms of normal modes.~\cite{Ashcroft76} We show that each normal mode is equipped with a characteristic relaxation timescale. Displacements along each normal mode produce stresses due to elastic deformation and internal dissipation, which are in force balance with loads due to external dissipation. Interestingly, the rheological behavior of a given normal mode can be described as a combination of a standard linear solid element due to elastic stresses and a Jeffreys model element due to the internal dissipative stresses. The rheological behavior of the system is then fully determined by connecting in parallel all the viscoelastic elements corresponding to individual normal modes. This behavior is generic and does not depend on the details of the microscopic model for dynamics. The method discussed in this paper shares similarities with several recent studies,~\cite{pessot2016dynamic, palyulin2018parameter, kriuchevskyi2020scaling} but it complements them by allowing treatment of various dissipative mechanisms.   

We applied the formalism to study the linear response of the vertex model,~\cite{honda1980much, farhadifar2007influence, fletcher2013implementing} widely used to describe the mechanical properties of epithelial tissues. Using the vertex model, Bi, et al. predicted that epithelial tissues can undergo rigidity transition at constant density by tuning cell properties.~\cite{bi2015density} These unexpected results have been observed in-vitro in human bronchial cell monolayers,~\cite{park2015unjamming} and have sparked interest in understanding how biological systems can take advantage of the presence of phase transitions and the accompanying tissue rheology.~\cite{petridou2019tissue,lennne2022sculpting} Understanding mechanical and rheological properties of the vertex model and the closely related self-propelled Voronoi model~\cite{bi2016motility, barton2017active, merkel2018geometrically} has, therefore, attracted significant attention.~\cite{moshe2018geometric, merkel2019minimal, tong2022linear, huang2022shear} Most studies to date~\cite{popovic2021inferring,das2021controlled, duclut2021nonlinear, huang2022shear} focused on the quasistatic regime including with large deformations, where plastic relaxation facilitated by cell rearrangements via T1 events leads to a strong non-linear response. Here, we take the opposite limit and explore the dynamic linear response of the vertex model over a wide range of driving frequencies and three different microscopic models of dissipation. In our previous work in Ref.~\cite{tong2022linear}, we numerically investigated the linear viscoelastic response of the vertex model with external dissipation. We found that even in the linear regime, the vertex model shows complex viscoelastic response, especially in the fluid phase, with multiple competing time scales. In this work, we demonstrated that the complex viscoelastic response of the vertex model can be accurately captured by normal modes. In addition, we show that internal and external dissipative mechanisms can result in a markedly different rheological response, emphasizing the importance of dissipation for the rheological behavior of model epithelial tissues.   

The paper is organized as follows. In Sec.~\ref{sec:model}, we use the normal mode analysis to derive the linear response rheological properties of overdamped soft materials with both external and internal dissipation. In Sec.~\ref{sec:vertex-model}, we apply this method to the vertex model of epithelial tissues and explicitly discuss three microscopic mechanisms of dissipation: 1)~external dissipation due to relative motion between a vertex and the substrate; 2)~internal dissipation due to relative motion of neighboring vertices, and 3)~internal dissipation due to the relative motion of neighboring cell centers. The first model is suitable for studies of cell monolayers supported by a solid substrate while the latter two are more appropriate for systems such as embryos. In Sec.~\ref{sec:discussion}, we provide concluding remarks. Details of the calculation of forces, the Hessian matrix, and stresses are presented in Appendices~\ref{app:force-on-vertex}-\ref{appendix:stress}. 

\section{Normal Modes Analysis}
\label{sec:model}
We model a soft viscoelastic system as a collection of $N$ interacting agents (e.g., vertices of a network, colloids in colloidal dispersion, cells in a tissue, etc.) whose interaction is described by a potential energy $E\left(\left\{\mathbf{R}_j\right\}\right)$, where $\{\mathbf{R}_j\}$ denotes the set of position vectors of all agents in a suitably chosen reference frame. The position vector of agent $j$ is $\mathbf{R}_j\in\mathbb{R}^d$ with $d$ being the number of spacial dimensions and $j \in \{1,\dots,N\}$. In general, the potential energy can be multi-body, and each agent $i$ experiences the force $-\nabla_{\mathbf{R}_i}E\left(\left\{\mathbf{R}_j\right\}\right)$. To simplify the notation, we introduce the column vector $\mathbf{r} \equiv \left(\mathbf{R}_1,\dots,\mathbf{R}_N\right)$, which contains positions of all agents, i.e., $\mathbf{r}\in\mathbb{R}^{dN}$. Similarly, a $dN$-dimensional vector, $-\nabla_{\mathbf{r}}E\left(\mathbf{r}\right)$, represents forces acting on all agents. Since typical systems of interests are in the overdamped regime, the relaxation dynamics of the system in the absence of an external driving force is described as
\begin{equation}\label{eq:motion}
    \hat{\boldsymbol{C}}\dot{\mathbf{r}}(t)=-\nabla_{\mathbf{r}}E\big(\mathbf{r}(t)\big),
\end{equation}
where the symmetric matrix $\hat{\boldsymbol{C}}$ provides a generalization of the single friction coefficient to more complex mechanisms of energy dissipation, which captures both external and internal dissipation. The overdot symbol is used to denote the time derivative of relevant quantities throughout this paper.

We are interested in the linear response of the system around an equilibrium state $\mathbf{r}^{\text{eq}}$, which corresponds to a solution of the set of algebraic equations,
\begin{equation}
    \nabla_{\mathbf{r}}E\big|_{\mathbf{r}=\mathbf{r}^\text{eq}}=0.
\end{equation}
In order to probe the linear response, we consider an external driving force $\mathbf{f}(t)$ that is sufficiently weak to produce deformations that are small compared to the typical distance between agents, i.e., it keeps the system in the basin of attraction of the local energy minimum. The corresponding displacements are $\delta\mathbf{r}(t)=\mathbf{r}(t)-\mathbf{r}^{\text{eq}}$ and linearized equations of motion take the form
\begin{equation}
    \hat{\boldsymbol{C}}\delta\dot{\mathbf{r}}(t)=-\hat{\boldsymbol{H}}\delta\mathbf{r}(t)+\mathbf{f}(t), \label{eq:motion_linear}
\end{equation}
where $\hat{\boldsymbol{H}}$ is the Hessian matrix with elements
\begin{equation}\label{eq:hessian}
    \hat{H}_{IJ} = \frac{\partial^2E}{\partial x_I\partial x_J}\Bigg|_{\mathbf{r}=\mathbf{r}^{\text{eq}}}.
\end{equation}
Here, $x_I$ and $x_J$ are, respectively, the $I^\text{th}$ and $J^\text{th}$ entries in the vector $\mathbf{r}$. Recall that $\hat{\boldsymbol{H}}$ is a real $dN\times dN$ symmetric matrix. If $\mathbf{r}^{\text{eq}}$ corresponds to a true local energy minimum, then $\hat{\boldsymbol{H}}$ is positive definite. Here, we relax this condition and assume that $\hat{\boldsymbol{H}}$ is positive semi-definite, i.e., we allow for the possibility that zero-energy modes are present in the system. Furthermore, in general, matrices $\hat{\boldsymbol{H}}$ and $\hat{\boldsymbol{C}}$ do not commute. 

In order to solve Eq.~(\ref{eq:motion_linear}), we consider the following generalized eigenvalue problem,~\cite{parlett1998symmetric}
\begin{equation}
    \hat{\boldsymbol{H}} \boldsymbol{\xi}_k = \lambda_k\hat{\boldsymbol{C}}\boldsymbol{\xi}_k,\label{eq:eigen_problem}
\end{equation}
where $\lambda_k$ and $\boldsymbol{\xi}_k$ are the $k^\text{th}$ eigenvalue and eigenvector, respectively. Since $\hat{\boldsymbol{H}}$ and $\hat{\boldsymbol{C}}$ are real symmetric matrices, the eigenvectors form a complete orthonormal basis $\mathcal{B}=\left\{\boldsymbol{\xi}_i|i=1,\dots,dN\right\}$, i.e., $\langle\boldsymbol{\xi}_i,\boldsymbol{\xi}_j\rangle_{\hat{\boldsymbol{C}}}=\delta_{ij}$, where $\delta_{ij}$ is the Kronecker delta and the inner product is defined as $\langle\mathbf{u},\mathbf{v}\rangle_{\hat{\boldsymbol{C}}} \equiv \mathbf{u}^\intercal\hat{\boldsymbol{C}}\mathbf{v}$. We can expand the displacement $\delta\mathbf{r}(t)$ in the basis $\mathcal{B}$ as 
\begin{equation}\label{eq:decomposition}
    \delta\mathbf{r}(t) = \sum_{k=1}^{dN} a_k(t) \boldsymbol{\xi}_k, 
\end{equation}
where $a_k(t)$ is the time-dependent amplitude of the normal mode $\boldsymbol{\xi}_k$. Equations of motion of the system in Eq.~\eqref{eq:motion_linear} can then be projected along each normal mode. Since normal modes are orthogonal to each other and, hence, decoupled, each behaves as an independent overdamped harmonic oscillator. The projection, therefore, leads to a set of decoupled dynamical equations for the amplitudes $a_k(t)$ of normal modes, 
\begin{equation}\label{eq:motion_mode}
    \dot{a}_k(t)=-\lambda_ka_k(t)+f_k(t),
\end{equation}
where $f_k(t)=\boldsymbol{\xi}_k^\intercal\mathbf{f}(t)$ is the projection of the driving force along the $k^\text{th}$ normal mode. We immediately see that $\lambda_k^{-1}$ is the characteristic relaxation time of the $k^\text{th}$ normal mode. For a given driving force $f_k$(t), Eq.~(\ref{eq:motion_mode}) can be solved as
\begin{equation}\label{eq:motion_integral} a_k(t)=a_k(0)e^{-\lambda_kt}+\int_0^tdt'f_k(t')e^{-\lambda_k(t-t')}.
\end{equation}
Thus, once the external driving force $\mathbf{f}(t)$ is specified, the dynamics of the system is fully determined by Eqs.~(\ref{eq:decomposition})-(\ref{eq:motion_integral}). Finally, Eq.~(\ref{eq:motion_mode}) has a simple solution in the frequency domain
\begin{equation} \label{eq:general_mode_coeff}
    \tilde {a}_k(\omega)=\frac{\tilde{f}_k(\omega)}{\lambda_k+i\omega},
\end{equation}
where the Fourier transform is defined as $\tilde {g}(\omega) = \int_{-\infty}^{+\infty} dt \, g(t) e^{-i \omega t}$.

\subsection{Application of normal modes to 2d systems with periodic boundaries under shear}

As an example of the general formalism presented above, we now demonstrate how to extract linear shear rheological properties of a $d=2$--dimensional system with periodic boundary conditions. Treating systems with open boundaries would be analogous, but one would have to solve the generalized eigenvalue problem in Eq.~(\ref{eq:eigen_problem}) subject to the appropriate boundary conditions. In numerical experiments, it is convenient to apply external driving via a macroscopic affine deformation of the entire system and measure the stress response in the system.~\cite{lerner2012unified,lerner2013simulations} One way of achieving this in a real system could be by placing the system on a sticky substrate and applying oscillatory shear deformation to the substrate. Due to interactions with the substrate, the deformation of the system follows the deformation of the substrate on short timescales, but the system can relax and produce non-affine motion on longer timescales. We remark that non-affine relaxation plays an important role in many soft systems, e.g., it is key for understanding the elastic properties of amorphous solids.~\cite{lemaitre2006sum}  

The affine deformation can be described as $\mathbf{R}_i(t)=\hat{\boldsymbol{F}}(t)\mathbf{R}^0_i$ where the initial position $\mathbf{R}^0_i\equiv\mathbf{R}_i(t=0)$ of agent $i$  is mapped to the current position $\mathbf{R}_i(t)$ by an affine deformation gradient tensor $\hat{\boldsymbol{F}}(t)$.~\cite{gurtin2010mechanics} For instance, the shear rheology of the system can be probed by applying an oscillatory affine simple shear described by the deformation gradient tensor
\begin{equation}\label{eq:F_simpleShear}
    \hat{\boldsymbol{F}}=\begin{pmatrix}1 & \epsilon(t)\\ 0 & 1\end{pmatrix},
\end{equation} 
where $\epsilon(t)=\epsilon_0 \sin(\omega_0 t)$ is the applied shear strain with amplitude $\epsilon_0\ll1$. The non-affine relaxation can be described by modifying the force balance in Eq.~(\ref{eq:motion}) as
\begin{equation}
\label{eq:motion_periodic}
    \hat{\boldsymbol{C}}\big(\dot{\mathbf{r}}(t)-\mathbf{v}^\text{aff}(t)\big)=-\nabla_{\mathbf{r}}E\big(\mathbf{r,\epsilon(t)}\big). 
\end{equation}
Here,  vector $\mathbf{v}^\text{aff}(t)$ contains the velocities of all agents due to the affine deformation imposed by the substrate, i.e., $\mathbf{v}^\text{aff}\equiv\left(\mathbf{V}_1^\text{aff},\dots,\mathbf{V}_N^\text{aff}\right)$, where $\mathbf{V}_i^\text{aff}(t)=\left(\frac{d}{dt}\hat{\boldsymbol{F}}(t)\right)\mathbf{R}_i^0$ is the affine part of the velocity of each individual agent. For the applied simple shear deformation in Eq.~(\ref{eq:F_simpleShear}), the affine velocity of agent $i$ is $\mathbf{V}_i^\text{aff}=\left(\dot{\epsilon}(t)R^{0,y}_{i},0\right)$, where $R^{0,y}_{i}$ is the $y-$coordinate of the agent $i$ in the initial position, and the periodic box is centered at the origin.

It is important to note that for a system with periodic boundary conditions, the energy, $E\left(\mathbf{r,\epsilon(t)}\right)$, also depends on the applied shear strain $\epsilon(t)$. This dependence enters via the $x$-component of the separation distance $R_{ij}^x$ between agents $i$ and $j$ as $R_{ij}^x=R_j^x-R_i^x+q_{ij}^x\ell_x+\epsilon(t)q_{ij}^y\ell_y$.~\cite{merkel2018geometrically} Here, $R_{i}^x$ and $R_{j}^x$ are the $x$-coordinates of the agents $i$ and $j$, respectively, $\ell_x$ and $\ell_y$ are the sizes of the rectangular simulation box in the $x$ and $y$ directions, respectively. $q_{ij}^y=0$ if agents $i$ and $j$ are connected without crossing the top or the bottom boundary and $q_{ij}^y=+1$ ($q_{ij}^y=-1$) if the bond connecting agent $i$ to agent $j$ crosses the top (bottom) boundary, with analogous expressions for $q_{ij}^x$ in terms of the left and right boundaries.~\cite{merkel2018geometrically}

Due to the dependence of the energy on the applied shear strain $\epsilon(t)$, the linearized equations of motion in Eq.~(\ref{eq:motion_periodic}) around the equilibrium state $\mathbf{r}^{\text{eq}}$ become 
\begin{equation}
     \hat{\boldsymbol{C}}\delta\dot{\mathbf{r}}(t)=-\hat{\boldsymbol{H}}\delta\mathbf{r}(t) + \overline{\mathbf{f}}^\text{pb} \epsilon(t) +\hat{\boldsymbol{C}}\mathbf{u}^\text{aff} \dot{\epsilon}(t).
    \label{eq:motion_linear_periodic}
\end{equation}
Here, we introduced the driving force $\overline{\mathbf{f}}^\text{pb}=-\frac{\partial^2E}{\partial {\mathbf{r}}\partial\epsilon}\Big|_{\mathbf{r}=\mathbf{r}^{\text{eq}},\epsilon=0}$ due to periodic boundaries and  rewrote the affine velocity as $\mathbf{v}^\text{aff}=\mathbf{u}^\text{aff} \dot{\epsilon}(t)$, where  $\mathbf{u}^\text{aff}=\left(\mathbf{U}_1^\text{aff},\dots,\mathbf{U}_N^\text{aff}\right)$ with $\mathbf{U}_i^\text{aff}=(R_i^{0,y},0)$. As discussed above, the displacements can be expanded in the basis of normal modes as in Eq.~(\ref{eq:decomposition}). In this basis, the equations of motions in Eq.~(\ref{eq:motion_linear_periodic}) become
\begin{equation}
    \dot{a}_k(t) = -\lambda_k a_k(t)+\alpha_k\epsilon(t)+\beta_k\dot{\epsilon}(t),
    \label{eq:motion_mode_periodic}
\end{equation}
where
\begin{subequations}
\label{eq:projected_coeffs}
\begin{alignat}{1}
\alpha_k&=\boldsymbol{\xi}_k^\intercal\,\overline{\mathbf{f}}^\text{pb}, \label{eq:alpha_k} \\
\beta_k&=\langle\boldsymbol{\xi}_k,\mathbf{u}^\text{aff}\rangle_{\hat{\boldsymbol{C}}}= \boldsymbol{\xi}_k^\intercal\hat{\boldsymbol{C}}\mathbf{u}^\text{aff}.\label{eq:beta_k}
\end{alignat}
\end{subequations}
The solution of Eq.~(\ref{eq:motion_mode_periodic}) in the frequency domain is 
\begin{equation} \label{eq:general_mode_coeff_periodic}
    \tilde {a}_k(\omega)=\left(\frac{\alpha_k + i \omega \beta_k}{\lambda_k+i\omega}\right)\, \tilde{\epsilon}(\omega).
\end{equation}

\begin{figure*}[t!]
    \centering
    \includegraphics{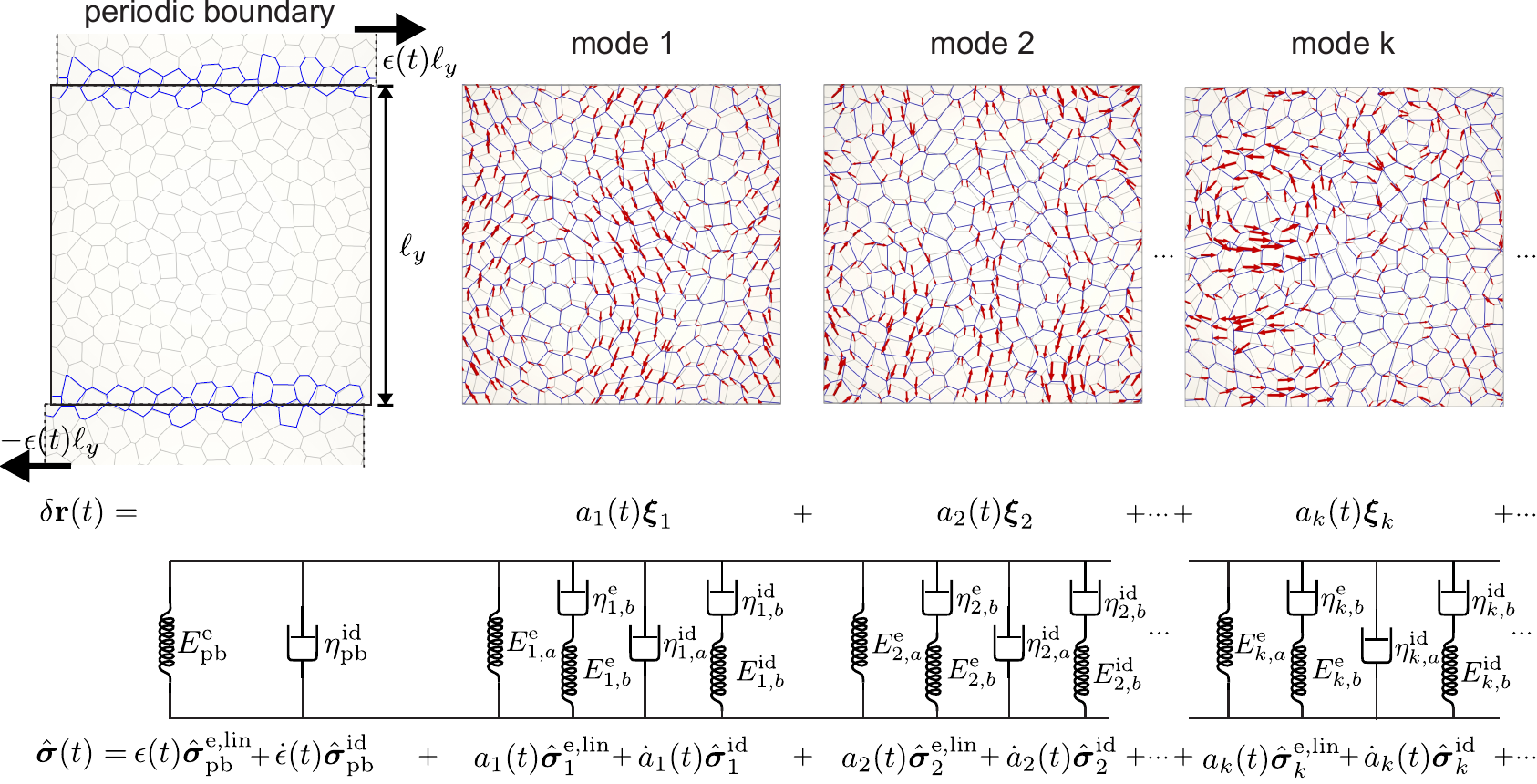}
    \caption{A schematic of the decomposition of the motion and the response stress along normal modes. The top left panel depicts the deformation of cells crossing the periodic boundary due to the relative motion of image boxes that are displaced according to the applied shear. The remaining panels show three representative normal modes for the vertex model. The grey mesh is the equilibrium configuration, the red arrows indicate displacements associated with the normal mode $\boldsymbol{\xi}_k$, and the blue mesh is the configuration after displacing vertices in the direction of the normal mode. Similarly, in the top left panel polygons outlined in blue indicate the distorted cells due to the relative movement of image boxes under applied shear.
    Perturbation from the equilibrium state $\delta\mathbf{r}(t)$ can be written as a linear superposition of displacements along the normal modes $\boldsymbol{\xi}_k$. The stress response of the system [see Eq.~(\ref{eq:total_stress})] due to shear deformation can be represented as a linear superposition of stresses $\epsilon(t) \hat{\boldsymbol{\sigma}}_\text{pb}^\text{e,lin}$ and $\dot{\epsilon}(t) \hat{\boldsymbol{\sigma}}_\text{pb}^\text{id}$ due to the elastic deformation and internal dissipation of cells crossing the periodic boundary, respectively, and stresses $a_k(t)\hat{\boldsymbol{\sigma}}_k^\text{e,lin}$ and $\dot{a}_k(t)\hat{\boldsymbol{\sigma}}^\text{id}_k$ due to the elastic deformation and internal dissipation of the $k^\text{th}$ normal mode, respectively. The rheological response of the system due to shear deformation [see Eq.~(\ref{eq:G_dynamics})] can thus be represented as a parallel sequence of a spring and a dashpot due to shear of cells crossing the periodic boundary and standard linear solid (SLS) model elements and Jeffreys model elements, where each SLS model element and Jeffreys model element describe the shear response of a normal mode. Expressions for spring constants and dashpot viscosities are given in the text. }
    \label{fig:modeSchematic}
\end{figure*}

\subsection{Stresses in 2d systems with periodic boundaries}

We measure the stress response of the system due to an external driving force as follows. The stress of the system has two contributions. One contribution is denoted as $\hat{\boldsymbol{\sigma}}^\text{e}(t)$, and  is generated by the interaction forces $-\nabla_{\mathbf{r}}E\left(\mathbf{r}\right)$ resulting from the elastic deformation. The other contribution is due to the internal dissipation resulting from the relative motion of agents with respect to each other, and it is denoted as $\hat{\boldsymbol{\sigma}}^\text{id}(t)$. Note that the loads $\mathbf{f}^{\text{ext}}(t)$ due to external dissipation induce stresses in the system via the force balance relation $\nabla \cdot (\hat{\boldsymbol{\sigma}}^\text{e}+\hat{\boldsymbol{\sigma}}^\text{id}) + \mathbf{f}^{\text{ext}} = 0$. In Appendix~\ref{appendix:stress}, we show a detailed derivation of these stress contributions for cells in the vertex model. Similar steps can be followed to derive the relevant expressions for stresses $\hat{\boldsymbol{\sigma}}^\text{e}$ and $\hat{\boldsymbol{\sigma}}^\text{id}$ for any other system.  

In the linear response regime, the displacement perturbed from the equilibrium state $\delta\mathbf{r}(t)$ can be written as a linear superposition of displacements along normal modes. The response stress $\hat{\boldsymbol{\sigma}}^\text{e}(t)$ due to the elastic deformation can thus be represented as a linear superposition of stress tensors $\hat{\boldsymbol{\sigma}}^\text{e}_k(t) = a_k(t) \hat{\boldsymbol{\sigma}}_k^\text{e,lin}$ due to the motion along each normal mode $k$ (see Fig.~\ref{fig:modeSchematic},  top right). The linear response tensor $\hat{\boldsymbol{\sigma}}_k^\text{e,lin}$ for $k^\text{th}$ normal mode can be calculated by measuring the response stress tensor $\delta\hat{\boldsymbol{\sigma}}^\text{e}_k$ due to perturbation $\delta\mathbf{r}=\delta a\boldsymbol{\xi}_k$ along mode $k$ so that $\hat{\boldsymbol{\sigma}}_k^\text{e,lin}=\delta\hat{\boldsymbol{\sigma}}^\text{e}_k/\delta a$, where the amplitude $\delta a$ is sufficiently small to produce deformations that are small compared to the typical distance between agents. There is an additional stress contribution due to the deformation of elastic bonds that cross the periodic boundary. This is because the displacements $\boldsymbol{\xi}_k$ of normal modes are periodic and they do not capture the relative displacement between the agent $i$ and its periodic image when the affine shear is applied. To account for that, we also consider stresses $\hat{\boldsymbol{\sigma}}^\text{e}_\text{pb}(t)=\epsilon(t) \hat{\boldsymbol{\sigma}}_\text{pb}^\text{e,lin}$ due to the deformation of the elastic bonds that cross periodic boundaries when the periodic images of the system are displaced according to the applied shear (see Fig.~\ref{fig:modeSchematic},  top left). The linear response tensor $\hat{\boldsymbol{\sigma}}_\text{pb}^\text{e,lin}$ can be calculated by applying a small shear $\delta \epsilon_0 \ll1$ to the periodic boundary, and measuring the resulting stress $\delta \hat{\boldsymbol{\sigma}}^\text{e}_\text{pb}$ so that $\hat{\boldsymbol{\sigma}}_\text{pb}^\text{e,lin}=\delta \hat{\boldsymbol{\sigma}}^\text{e}_\text{pb}/\delta \epsilon_0$. 
Therefore, the total elastic stress tensor is
\begin{equation}\label{eq:stress}
    \hat{\boldsymbol{\sigma}}^\text{e}(t)=\epsilon(t) \hat{\boldsymbol{\sigma}}_\text{pb}^\text{e,lin} + \sum_k a_k(t)\hat{\boldsymbol{\sigma}}_k^\text{e,lin}.
\end{equation}

The stresses $\hat{\boldsymbol{\sigma}}^\text{id}(t)$ generated by the internal dissipation, i.e., the motion of agents relative to each other, are proportional to the velocities of the agents. The velocity is expanded in the basis $\mathcal{B}$ as $\delta\dot{\mathbf{r}}(t)=\sum_k \dot{a}_k(t)\boldsymbol{\xi}_k$.
The stress tensor $\hat{\boldsymbol{\sigma}}^\text{id}(t)$ can thus be written as a linear superposition,
\begin{equation}
 \hat{\boldsymbol{\sigma}}^\text{id}(t)=\dot{\epsilon}(t) \hat{\boldsymbol{\sigma}}_\text{pb}^\text{id} + \sum_k \dot{a}_k(t)\hat{\boldsymbol{\sigma}}^\text{id}_k,   
\end{equation}
where $\dot{\epsilon}(t) \hat{\boldsymbol{\sigma}}_\text{pb}^\text{id}$ is the dissipative stress due to the relative motion of agents when the periodic images of the system are displaced according to the applied shear, and $\hat{\boldsymbol{\sigma}}^\text{id}_k$ is the dissipative stress generated by the $k^\text{th}$ normal mode. Note that the stresses due to internal dissipation are linear by definition because the dissipative forces depend linearly on the agent velocities [see Eq.~(\ref{eq:motion})]. Then the total stress is 
\begin{eqnarray}
\hat{\boldsymbol{\sigma}}(t)&=&\hat{\boldsymbol{\sigma}}^\text{e}(t) + \hat{\boldsymbol{\sigma}}^\text{id}(t), \nonumber \\
\hat{\boldsymbol{\sigma}}(t)&=&\epsilon(t) \hat{\boldsymbol{\sigma}}_\text{pb}^\text{e,lin} + \sum_k a_k(t)\hat{\boldsymbol{\sigma}}_k^\text{e,lin}\nonumber\\ &&+\dot{\epsilon}(t) \hat{\boldsymbol{\sigma}}_\text{pb}^\text{id} +  \sum_k \dot{a}_k(t)\hat{\boldsymbol{\sigma}}^\text{id}_k.
\label{eq:total_stress}
\end{eqnarray}
\subsection{Storage and loss moduli in 2d systems with periodic boundaries}

For the shear rheology, the dynamic shear modulus is defined as,~\cite{larson1999structure}
\begin{equation}
\begin{split}
\label{eq:dynamic_shear_modulus}
G^*(\omega)=\frac{\tilde{\hat{\sigma}}_{xy}(\omega)}{\tilde{\epsilon}(\omega)} =& \quad \quad G^\text{e}_\text{pb} \  + \sum_k  \left(\frac{\alpha_k + i \omega \beta_k}{\lambda_k+i\omega}\right) G_k^\text{e}\\
&+i\omega G^\text{id}_\text{pb}+ \sum_k i\omega  \left(\frac{\alpha_k + i \omega \beta_k}{\lambda_k+i\omega}\right) G_k^\text{id},
\end{split}
\end{equation}
where $G^\text{e}_\text{pb}\equiv \hat{{\sigma}}_{\text{pb},xy}^\text{e,lin}$ and $G^\text{id}_\text{pb}\equiv \hat{{\sigma}}_{\text{pb},xy}^\text{id}$ are the moduli due to the shear of the periodic boundary and
\begin{subequations}
\label{eq:projected_stress}
\begin{align}
    G_k^\text{e}& \equiv \hat{{\sigma}}_{k,xy}^\text{lin},\label{eq:G_e}\\
    G_k^\text{id}& \equiv \hat{{\sigma}}_{k,xy}^\text{id}.\label{eq:G_id}
\end{align}
\end{subequations}
Note that $G^\text{id}_\text{pb}$, $G_k^\text{e}$, and $G_k^\text{id}$ themselves do not have units of stress. We, however, use the letter ``G'' to emphasize their contributions to the dynamic moduli, which have units of stress.
In general, the dynamic shear modulus $G^*(\omega)=G^\prime(\omega)+iG^{\prime\prime}(\omega)$ has a real part $G^\prime$, which is called the storage modulus, and an imaginary part $G^{\prime\prime}$, which is called the loss modulus.~\cite{larson1999structure}
Using Eq.~(\ref{eq:dynamic_shear_modulus}), one can write the storage and loss moduli as
\begin{subequations}
\label{eq:G_dynamics}
\begin{equation}
\begin{split}
    G^\prime(\omega)=&G^\text{e}_\text{pb} +\sum_k \left(\frac{\alpha_k \lambda_k + \beta_k\omega^2}{\lambda_k^2+\omega^2}\right)G_k^\text{e}\\
    &-\sum_k  \left(\frac{(-\alpha_k+\beta_k \lambda_k)\omega^2}{\lambda_k^2+\omega^2}\right)G_k^\text{id},\\ \label{eq:G_prime}
\end{split}
\end{equation}
\begin{equation}
\begin{split}
G^{\prime\prime}(\omega)=&\omega G^\text{id}_\text{pb}+\sum_k  \left(\frac{(-\alpha_k+\beta_k \lambda_k)\omega}{\lambda_k^2+\omega^2}\right)G_k^\text{e}\\
&+\sum_k \left(\frac{(\alpha_k \lambda_k + \beta_k\omega^2)\omega}{\lambda_k^2+\omega^2}\right)G_k^\text{id}. \label{eq:G_double_prime}
\end{split}
\end{equation}
\end{subequations}

Upon a closer inspection of the expressions for the storage and loss moduli, we recognize that the contributions due to periodic boundaries can be represented as a Kelvin-Voigt element and that each normal mode $k$ behaves as a standard linear solid element connected in parallel with a Jeffreys element.~\cite{larson1999structure} The Kelvin-Voigt element can be represented as a spring  $E^\text{e}_\text{pb}=G^\text{e}_\text{pb}$ connected in parallel with a dashpot
$\eta^\text{id}_\text{pb}=G^\text{id}_\text{pb}$. 
The standard linear solid element can be represented as a spring with elastic constant $E_{k,a}^\text{e}=G_k^\text{e} \alpha_k/\lambda_k$ connected in parallel with a Maxwell element that consists of a spring $E_{k,b}^{\text{e}}=G_k^\text{e} (-\alpha_k+\beta_k \lambda_k)/\lambda_k$ and a dashpot $\eta_{k,b}^{\text{e}} = G_k^\text{e} (-\alpha_k+\beta_k \lambda_k)/\lambda_k^2$ connected in series (see Fig.~\ref{fig:modeSchematic}). The Jeffreys model element can be represented as a dashpot $\eta_{k,a}^\text{id}=G_k^\text{id}\beta_k$ connected in parallel with a Maxwell element that consists of a spring $E_{k,b}^{\text{id}}=G_k^\text{id}(\alpha_k-\beta_k\lambda_k)$ and a dashpot $\eta_{k,b}^{\text{id}}=G_k^\text{id}(\alpha_k-\beta_k\lambda_k)/\lambda_k$ (see Fig.~\ref{fig:modeSchematic}). The characteristic time scale for mode $k$ is, therefore, $\eta_{k,b}^{\text{e}}/E_{k,b}^{\text{e}}=\eta_{k,b}^{\text{id}}/E_{k,b}^{\text{id}}=1/\lambda_k$. The full rheological response of the system can thus be represented as a Kelvin-Voigt element
due to periodic boundary and a sequence of standard linear solid elements and Jeffreys model elements connected in parallel, each standard linear solid element and Jeffreys model element corresponding to the contribution from one normal mode (see Fig.~\ref{fig:modeSchematic}). The dynamic response of the system is, therefore,  characterized by a spectrum of relaxation timescales $\lambda_k^{-1}$ corresponding to each normal mode.

Finally, we note that the contribution to the loss modulus due to internal dissipation scales as $G^{\prime\prime}(\omega)\sim\omega$ at high frequencies [see Eq.~(\ref{eq:G_double_prime})]. This is in contrast to the case with external dissipation only, where the loss modulus scales as $G^{\prime\prime}(\omega) \sim 1/\omega$ [see Eq.~(\ref{eq:G_double_prime})].
Thus, in general, at high frequencies, the loss modulus increases linearly with frequency and is dominated by internal dissipation. 
As we show below, the onset of the crossover to the linear scaling of the loss modulus with $\omega$ in systems with internal dissipation, however, depends on the values of parameters and can start to occur at relatively high values of $\omega$, revealing rich low-frequency behavior. 

\begin{figure*}[t!]
    \centering
    \includegraphics{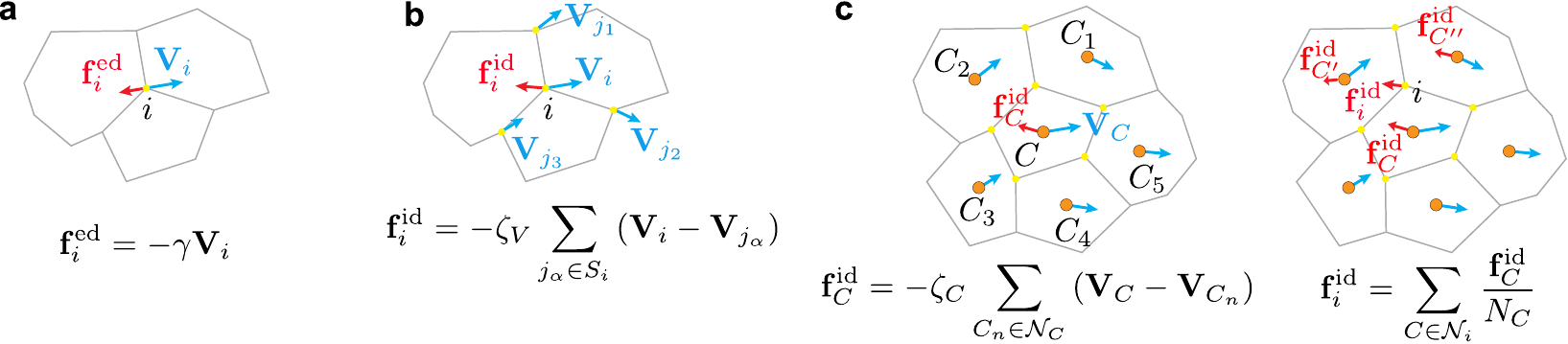}
    \caption{Schematics of the three types of dissipation mechanisms in the vertex model considered in this paper. Velocity vectors are shown in blue and friction forces are red. (a)~Dissipation due to the relative motion between vertices and a solid substrate. Vertex $i$ experiences a frictional force proportional to its velocity $\mathbf{V}_i$, with friction coefficient $\gamma$. (b)~Friction force on the vertex $i$ is due to its motion relative to neighboring vertices. $\mathcal{S}_i$ is the set of all vertices connected to vertex $i$ by a cell-cell junction. (c)~Dissipation is due to the relative motion of neighboring cell centers. $\mathcal{N}_C$ includes all neighboring cells of cell $C$. $\mathcal{N}_i$ includes all cells that share vertex $i$. $\mathbf{f}_C^\text{id}$ is the friction force that cell $C$ experiences due to relative motion with respect to its neighboring cells. $\mathbf{f}_i^\text{ed}$ and $\mathbf{f}_i^\text{id}$ are the total friction forces applied at vertex $i$ due to external and internal dissipation, respectively. $N_C$ is the number of vertices that belong to cell $C$.}
    \label{fig:frictionSchematic}
\end{figure*}

\section{Application to the Vertex model with different types of dissipation}
\label{sec:vertex-model}
In this section, we apply this formalism to analyze the linear viscoelastic properties of the vertex model of epithelial tissues with both external (i.e., cell-substrate) and internal (i.e., cell-cell) dissipation. The results are compared against direct numerical simulations of the vertex model. In the vertex model, a confluent epithelial tissue is represented as a polygonal tiling of the plane. We also assume that the tiling is subject to periodic boundary conditions. Two cells share a junction, which is modeled as a straight line, and three or more cells meet at a vertex, which is the degree of freedom in the model. The mechanical energy of the system is, 
\begin{equation}
    E=\sum_{C}\left[\frac{K}{2}\left(A_C-A_{0}\right)^2+\frac{\Gamma}{2}\left(P_C-P_{0}\right)^2\right], \label{eq:vm_energy}
\end{equation}
where $K$ and $\Gamma$ are the area and perimeter moduli, and $A_C$ and $P_{C}$ are the area and perimeter of cell $C$, respectively. $A_0$ and $P_0$ are, respectively, the preferred area and perimeters, here assumed  to be the same for all cells. The dimensionless cell-shape parameter $p_0=P_0/\sqrt{A_0}$ controls whether the model tissue is in the solid or the fluid regime.~\cite{bi2015density} Model tissues with low (high) values of the cell-shape parameters $p_0<p_c$ ($p_0>p_c$) behave like solids (fluids). The critical value of the cell-shape parameter $p_c$ that characterizes the solid-fluid transition is $p_c = \sqrt{8\sqrt{3}} \approx 3.722$ for regular hexagonal tilings,~\cite{staple2010mechanics} while for disordered tilings the critical value falls in the range $p_c \approx 3.8 - 3.9$~\cite{bi2015density,merkel2019minimal,wang2020anisotropy} and depends on the procedure with which they are generated. Here, we consider both regular hexagonal and disordered tilings. Expressions for mechanical forces on each vertex $-\nabla_{\mathbf{r}}E$, the Hessian matrix $\hat{\boldsymbol{H}}$, and the driving force $\overline{\mathbf{f}}^\text{pb}$ due to periodic boundaries that are used for the normal modes analysis are given in Appendices~\ref{app:force-on-vertex}, \ref{appendix:Hessian}, and \ref{appendix:driving_fb}, respectively.

The dynamics of the vertex model has been studied almost exclusively with the assumption that the only source of dissipation is the interaction between the vertex and the substrate.~\cite{fletcher2014vertex} In real epithelial tissues, however, there are different sources of dissipation, many of which have not been well understood. We, therefore, studied three simple dissipation mechanisms, as shown in Fig.~\ref{fig:frictionSchematic}. These are encoded in the matrix $\hat{\boldsymbol{C}}$ and describe the dynamics of vertices [see Eq.~(\ref{eq:motion})]. Specifically, Fig.~\ref{fig:frictionSchematic}a shows the friction $\mathbf{f}_i^\text{ed} = -\gamma \mathbf{V}_i$ due to the relative velocity $\mathbf{V}_i$ between the vertex $i$ and the substrate. The superscript `ed' is used to emphasize that this is a source of external dissipation.
In this case, the dissipation matrix takes a simple diagonal form, $\hat{\boldsymbol{C}}=\gamma\hat{\boldsymbol{I}}$, where $\gamma$ is the friction coefficient and $\hat{\boldsymbol{I}}$ is the identity matrix.
In Fig.~\ref{fig:frictionSchematic}b, we consider the internal dissipation due to relative motions of neighboring vertices with friction coefficient $\zeta_V$. In this case, the dissipation force for the vertex $i$ can be expressed as $\mathbf{f}_i^\text{id} = -\zeta_V \sum_{j_\alpha \in S_i} \left( \mathbf{V}_i -\mathbf{V}_{j_\alpha} \right)$, where the summation is over all nearest neighbor vertices, referred to as the ``star'' of vertex $i$ and denoted as $\mathcal{S}_i$. Assuming that each vertex is shared by three cell-cell junctions, $3\zeta_V$ appears on the diagonal of the $\hat{\boldsymbol{C}}$ matrix, and each row has three nonzero off-diagonal elements with the value $-\zeta_V$. Clearly, if there are vertices with coordination higher than three, this will be reflected in the diagonal term and the row of the matrix $\hat{\boldsymbol{C}}$ that correspond to such vertices. Another formulation of the internal dissipation is to consider the relative motions of neighboring geometric centers of cells, as shown in Fig.~\ref{fig:frictionSchematic}c. The velocity $\mathbf{V}_C$ of the geometric center of cell $C$ is defined as the average velocity of the $N_C$ vertices that belong to it, i.e., $\mathbf{V}_C=\frac{1}{N_C}\sum_{i \in C}\mathbf{V}_i$. The friction force that cell $C$ experiences due to the relative motions with respect to its neighboring cells is then defined as $\mathbf{f}_C^\text{id}=-\zeta_C\sum_{C_n\in\mathcal{N}_C}\left(\mathbf{V}_C-\mathbf{V}_{C_n}\right)$, where $\zeta_C$ is the friction coefficient and  the set $\mathcal{N}_C$ includes all neighboring cells of cell $C$. This friction force is assumed to be equally distributed across all the $N_C$ vertices that belong to cell $C$. Thus, the total friction force on the vertex $i$ is denoted as $\mathbf{f}_i^\text{id}=\sum_{C\in\mathcal{N}_i}\mathbf{f}_C^\text{id}/N_C$, where the set $\mathcal{N}_i$ includes all cells that share vertex $i$. Once the friction force at each vertex is known, the dissipation matrix $\hat{\boldsymbol{C}}$ can be written accordingly. Note that in this dissipation model, more than three vertices contribute to the force on a given vertex each contributing to an off-diagonal element of the matrix $\hat{\boldsymbol{C}}$. We also note that the models characterizing internal dissipation, such as the ones in Fig.~\ref{fig:frictionSchematic}b, c, play an important role if cells are not supported by the substrate, e.g., as is the case in early-stage embryos. Finally, we remark that in the absence of external dissipation the $\hat{\boldsymbol{C}}$ matrix is singular reflecting the fact that translations and rigid body rotations do not cause internal dissipation. Therefore, one should either consider small, but finite external friction, which is done in this study, or include inertia.

In the remainder of this section, we show how those three different types of dissipation affect the rheological properties of the vertex model for regular hexagonal and disordered tilings. For all three models, we compared the results of the normal mode analysis with direct numerical simulations of the vertex model.

\subsection{Simulation setup}
\begin{figure*}
    \centering
    \includegraphics{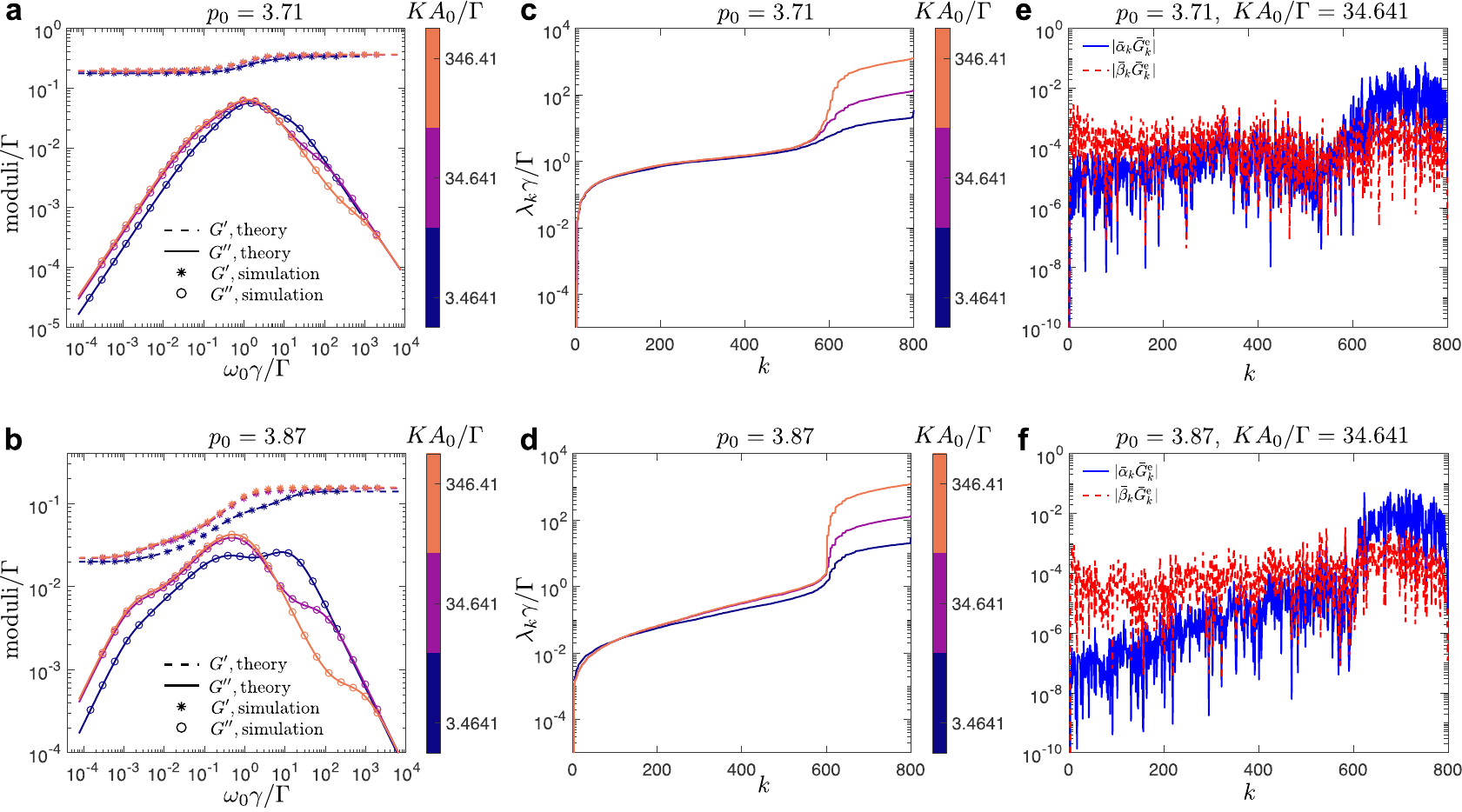}
    \caption{Shear rheology of disordered tilings with dissipation due to friction with the substrate for different values of the area elastic moduli $K$. Results for two representative values of the cell shape parameter, $p_0=3.71$ (top row) and $p_0=3.87$ (bottom row), are shown. (a,b)~Storage and loss moduli from the simulations~(symbols) compared with the predictions of the normal mode analysis~(lines). Different colors represent the results from different values of the ratio $KA_0/\Gamma$. (c,d)~Nonzero eigenvalues $\lambda_k$ in ascending order for different values of $KA_0/\Gamma$. (e,f)~Absolute values of normalized coefficients $|\Bar{\alpha}_k\Bar{G}^\text{e}_k|=|\alpha_kG_k^\text{e}\gamma/(KA_0)^2|$ and $|\Bar{\beta}_k\bar{G}^\text{e}_k|=|\beta_kG_k^\text{e}/(KA_0)|$ at one representative value of $KA_0/\Gamma=34.641$.}
    \label{fig:modulus_areaTerm}
\end{figure*}

We start by briefly summarizing the setup of our vertex model simulations, with additional details provided in Ref.~\cite{tong2022linear}. In most simulations of the vertex model [Eq.~(\ref{eq:vm_energy})], we fixed the values of $K$ and $A_0$, and measured the energy in units $KA_0^2$, stresses in units $KA_0$, and lengths in unit $A_0^{1/2}$. The preferred cell perimeter $P_0$, was varied to tune the system between solid or fluid phases. We fixed the perimeter modulus $\Gamma$ by fixing the ratio $KA_0/\Gamma\approx3.464$ in most simulations since it does not affect the location of the solid to fluid transition. In Fig.~\ref{fig:modulus_areaTerm} we, however, show an example of how the ratio $KA_0/\Gamma$ affects the dynamic shear modulus by changing the area modulus $K$.

We created regular hexagonal as well as disordered tilings subject to periodic boundary conditions as described in Ref.~\cite{tong2022linear}. All configurations used to probe the rheology corresponded to local energy minima obtained by the FIRE minimization algorithm.~\cite{bitzek2006structural} The shear rheology was probed by applying an oscillatory affine simple shear described by the deformation gradient $\hat{\boldsymbol{F}}=\big(\begin{smallmatrix}1 & \epsilon(t)\\ 0 & 1\end{smallmatrix}\big)$, where $\epsilon\left(t\right)=\epsilon_0\sin\left(\omega_0 t\right)$ and we used a small magnitude of deformation, i.e., $\epsilon=10^{-7}$. At each time step, we first applied the affine shear deformation to the simulation box and all vertices, which was followed by internal relaxation of vertices according to the overdamped dynamics [see Eq.~(\ref{eq:motion})].

\begin{figure*}[t!]
    \centering
    \includegraphics{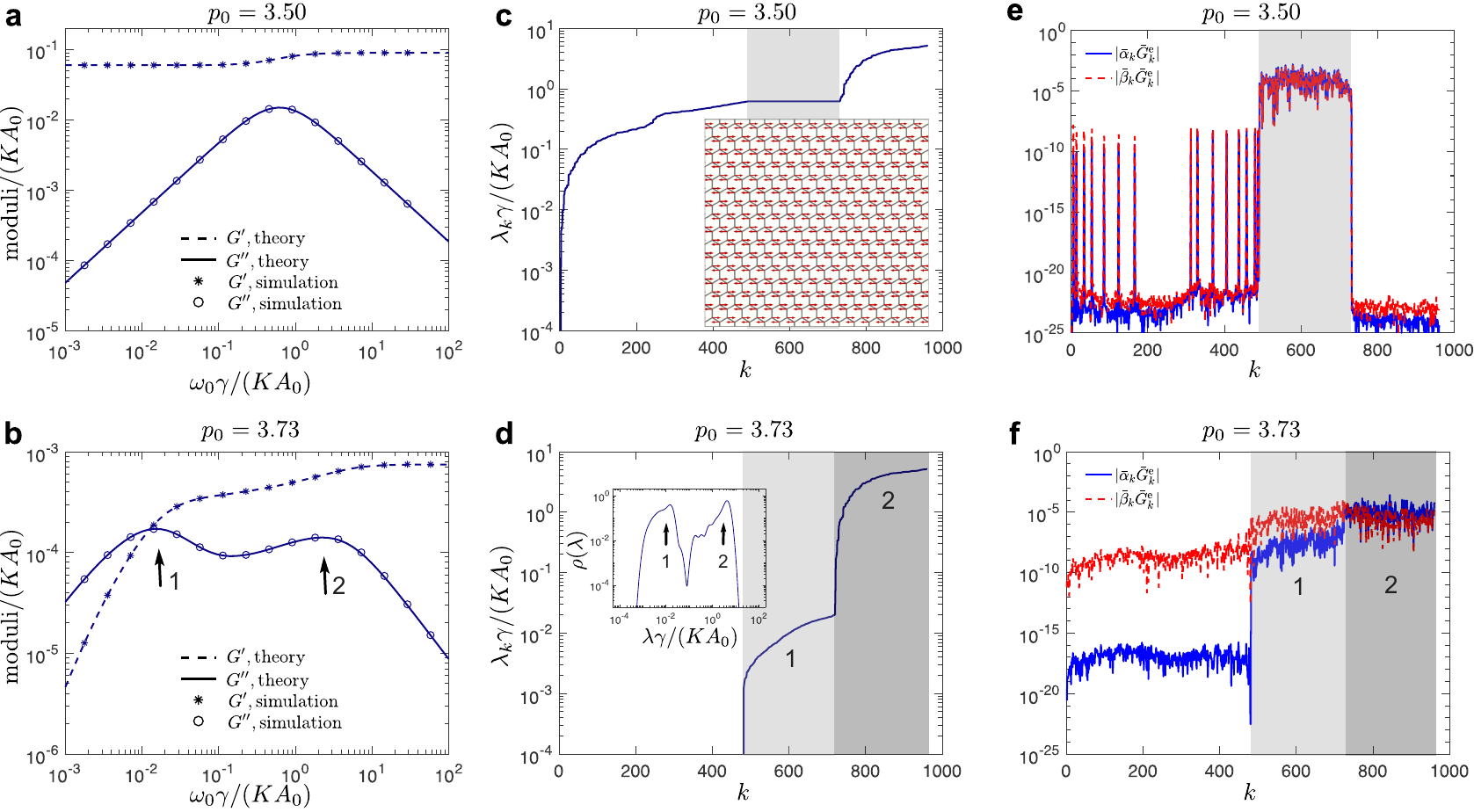}
    \caption{Shear rheology of hexagonal tilings with dissipation due to friction with the solid substrate. Results for two representative values of the cell shape parameter, $p_0=3.5$ in the solid phase (top row) and $p_0=3.73$ in the fluid phase (bottom row) are shown. (a,b)~Storage and loss moduli from the simulations~(symbols) compared with the predictions based on the normal mode analysis~(lines). (c,d)~Nonzero eigenvalues $\lambda_k$ vs.~the mode number $k$; eigenvalues are sorted in the ascending order. (e,f)~Absolute values of normalized coefficients $|\Bar{\alpha}_k\Bar{G}^\text{e}_k|=|\alpha_k G_k^\text{e}\gamma/(KA_0)^2|$ and $|\Bar{\beta}_k\bar{G}^\text{e}_k|=|\beta_k G_k^\text{e}/(KA_0)|$ [see Eqs.~(\ref{eq:projected_coeffs}) and (\ref{eq:projected_stress})]. Also note that the ordinate covers $\sim25$ decades. The inset in (c)~shows a schematic of the normal mode $\boldsymbol{\xi}_D$ that dominates the shear rheology of hexagonal tilings in the solid phase. This dominant normal mode is a linear combination of degenerate normal modes within the shaded region in panel (c,e). Labeled arrows in panel (b) denote peaks that correspond to the characteristic timescales from the normal modes in the shaded regions in panels (d) and (f). The inset in (d) shows the density of states $\rho(\lambda)$.}
    \label{fig:moduli_honeycomb}
\end{figure*}

The response stress tensor, $\hat{\boldsymbol{\sigma}}_C\left(t\right)=\hat{\boldsymbol{\sigma}}^\text{e}_C\left(t\right)+\hat{\boldsymbol{\sigma}}^\text{id}_C\left(t\right)$, for each cell $C$ was computed following the formalism introduced in Refs.~\cite{chiou2012mechanical,yang2017correlating,nestor2018relating,jensen2022couple} but took into account the contribution from internal dissipative forces (see Appendix~\ref{appendix:stress} for details). The stress due to elastic deformation is 
\begin{equation}
\hat{\boldsymbol{\sigma}}^\text{e}_C=-\Pi_C\hat{\boldsymbol{I}}+\frac{1}{2A_C}\sum_{e\in C}\mathbf{T}_e\otimes\mathbf{l}_e,
\label{eq:cell_stress_elastic}
\end{equation} where the summation is over all junctions $e$ belonging to cell $C$. In the above Eq.~\eqref{eq:cell_stress_elastic}, $\Pi_C = -\frac{\partial E}{\partial A_C}=-K\left(A_C-A_0\right)$ is the hydrostatic pressure inside cell $C$, $\hat{\boldsymbol{I}}$ is the unit tensor, and $\mathbf{T}_e=\frac{\partial E}{\partial\mathbf{l}_e}=2 \Gamma\left(P_C-P_0\right) \mathbf{l}_e/|\mathbf{l}_e|$ is the tension along the junction $e$ with $\mathbf{l}_e$ being a vector joining the two vertices on it.~\cite{chiou2012mechanical,yang2017correlating,nestor2018relating}
The stress due to internal dissipation is 
\begin{equation}
  \hat{\boldsymbol{\sigma}}^\text{id}_C = -\frac{1}{2z_i A_{C}}\sum_{i\in C}\left(\tilde{\mathbf{R}}_{i}\otimes\mathbf{f}_{i}^{\text{id}}+\mathbf{f}_{i}^{\text{id}}\otimes\tilde{\mathbf{R}}_{i}\right), \label{eq:stress_id} 
\end{equation}
where the summation is over all vertices $i$ belonging to cell $C$.  In Eq.~(\ref{eq:stress_id}), $\mathbf{f}_{i}^{\text{id}}$ is the internal friction force applied at vertex $i$, and $\tilde{\mathbf{R}}_{i}=\mathbf{R}_{i}-\mathbf{R}_{C}$ is the position of vertex $i$ relative to the cell's geometric center $\mathbf{R}_{C} = \sum_{i\in C} \mathbf{R}_{i}/N_C$. The average stress tensor $\hat{\boldsymbol{\sigma}}\left(t\right)=\sum_C w_C\hat{\boldsymbol{\sigma}}_C\left(t\right)$, with $w_C = A_C/\sum_C A_C$, was used as a measure for the response of the system. We recorded the average shear stress signal $\hat{{\sigma}}_{xy}\left(t\right)$ once the system reached a steady state. The dynamics modulus $G^*(\omega_0)=\tilde{\hat{{\sigma}}}_{xy}(\omega_0)/\tilde{\epsilon}(\omega_0)$ was computed at a given driving frequency $\omega_0$ of the applied strain, where $\tilde{\hat{{\sigma}}}_{xy}(\omega)$ and $\tilde{\epsilon}(\omega)$ are the Fourier transforms of $\hat{{\sigma}}_{xy}(t)$ and $\epsilon(t)$, respectively.

\subsection{Friction between vertices and a substrate}
We first performed simulations of the vertex model with external dissipation only, i.e., the friction force $-\gamma\mathbf{V}_i$ applied to each vertex, as shown in Fig.~\ref{fig:frictionSchematic}a. Since the external friction force produce stresses only indirectly (see Appendix~\ref{appendix:stress} for details), the values of $G^{\text{id}}_\text{pb}$ and $G^{\text{id}}_k$ are $0$ in Eq.~(\ref{eq:G_dynamics}). We validated the normal mode approach by comparing the shear moduli with the results of simple shear simulations of regular hexagonal tilings, which have been analyzed extensively in our previous work~\cite{tong2022linear}. Fig.~\ref{fig:moduli_honeycomb}a,b show the storage and loss moduli obtained from simulations (dots) for representative values of $p_0$ in (a)~the solid and (b)~the fluid phase, which show excellent agreement with the predictions (lines) of the normal mode analysis. The corresponding eigenvalues $\lambda_k$ and the normalized coefficients $\Bar{\alpha}_k\Bar{G}^\text{e}_k=\alpha_kG_k^\text{e}\gamma/(KA_0)^2$ and $\Bar{\beta}_k\bar{G}^\text{e}_k=\beta_kG_k/(KA_0)$ [see Eq.~(\ref{eq:projected_coeffs}), Eq.~(\ref{eq:projected_stress}) and Eq.~(\ref{eq:G_dynamics})] for the normal mode analysis are shown in Figs.~\ref{fig:moduli_honeycomb}c,d and \ref{fig:moduli_honeycomb}e,f, respectively. 

We note that the distribution of eigenvalues $\lambda_k$ is often presented in the form of the density of states \cite{lerner2012unified,palyulin2018parameter,kriuchevskyi2020scaling}. For the present analysis, however, in addition to the eigenvalues one also needs to know the values of coefficients $\Bar{\alpha}_k\Bar{G}^\text{e}_k$ and $\Bar{\beta}_k\bar{G}^\text{e}_k$ that describe projections of external driving force to the normal modes. For this reason, it is more instructive to present $\lambda_k$, $\Bar{\alpha}_k\Bar{G}^\text{e}_k$, and $\Bar{\beta}_k\bar{G}^\text{e}_k$ as functions of $k$. For reference, however, we show the density of states in the inset in Fig.~\ref{fig:moduli_honeycomb}d.

In our previous work~\cite{tong2022linear}, we noted that the rheological response of regular hexagonal tilings in the solid and fluid phases can be well described with the standard linear solid and Burgers model, respectively. This can be explained with the help of normal modes. The shear response in the solid phase is dominated by the set of normal modes in the shaded region in Fig.~\ref{fig:moduli_honeycomb}e, where coefficients $\Bar{\alpha}_k\Bar{G}^\text{e}_k$ and $\Bar{\beta}_k\bar{G}^\text{e}_k$ have the highest values. All modes in this region correspond to the identical eigenvalue that we denote as $\lambda_D$ (see the shaded region in Fig.~\ref{fig:moduli_honeycomb}c). Hence, the shear response of hexagonal tilings in the solid phase is characterized by a single time scale $\lambda_D^{-1}$, and, therefore, can be accurately captured by the standard linear solid model. Note that the response of the system is dominated by the linear combination of normal modes that corresponds to the projection of the affine shear deformation to the set of these degenerate normal modes as 
 $\boldsymbol{\xi}_D=\sum_{k\text{ s.t. }\lambda_k=\lambda_D} \left(\boldsymbol{\xi}_k^\intercal\,\mathbf{u}^\text{aff} \right)\,\boldsymbol{\xi}_k$ of the same eigenvalue $\lambda_D$.  
 The normal mode $\boldsymbol{\xi}_D$ is shown in the inset of Fig.~\ref{fig:moduli_honeycomb}b, where it is  represented by the corresponding displacement field, which shows that the connected vertices move horizontally by the same amount but in opposite directions.

\begin{figure*}[t!]
    \centering
    \includegraphics{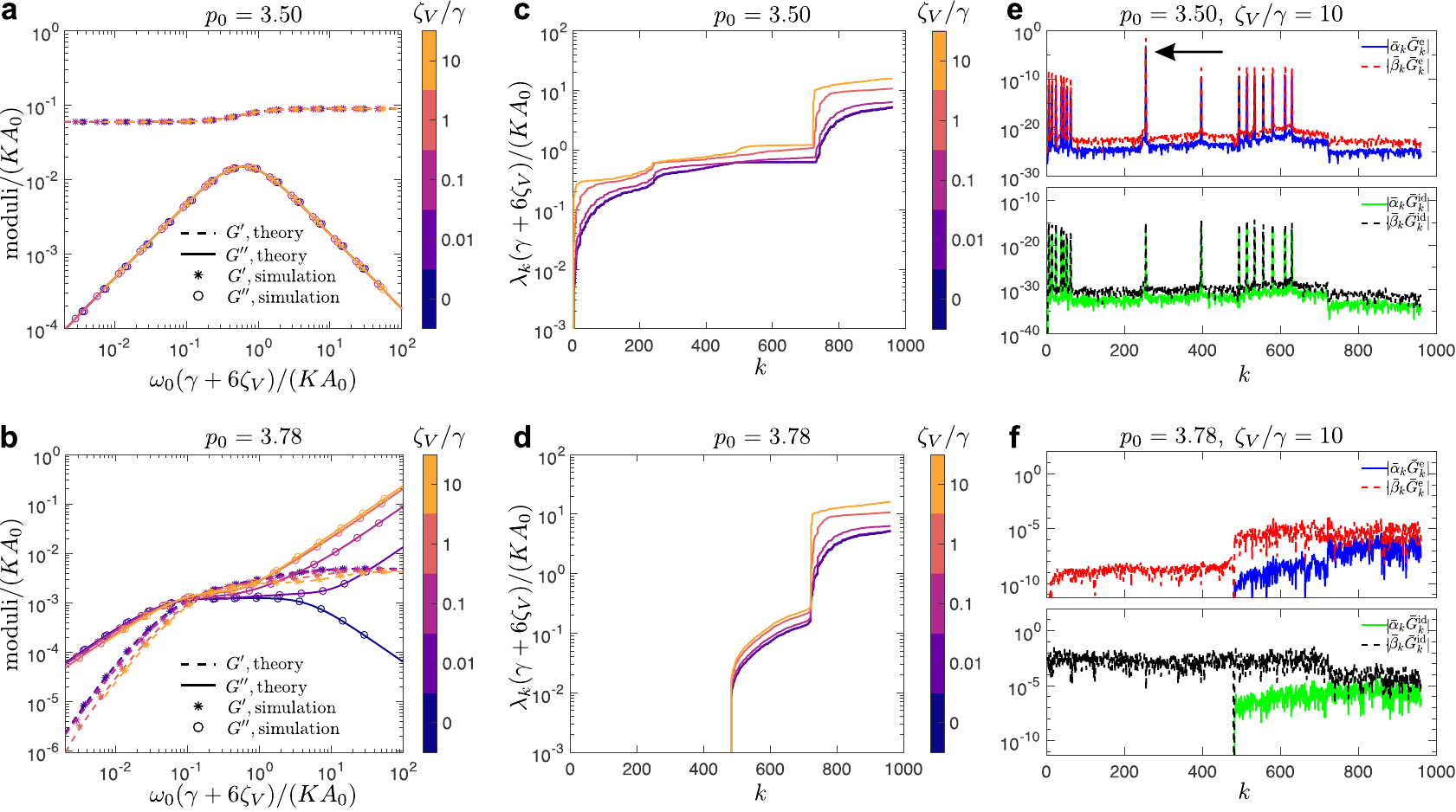}
    \caption{Shear rheology of hexagonal tilings with internal dissipation due to the relative motion of neighboring vertices in addition to the vertex-substrate friction. Results for two representative values of the cell shape parameter, $p_0=3.5$ in the solid phase (top row) and $p_0=3.78$ in the fluid phase (bottom row) are shown. (a,b)~Storage and loss moduli from the simulations (symbols) compared with the predictions from normal modes (lines) for different values of the internal friction coefficient $\zeta_V$ (see colorbar). 
    (c,d)~Nonzero eigenvalues $\lambda_k$ in ascending order for different values of $\zeta_V$. (e,f)~Normalized coefficients $\Bar{\alpha}_k\Bar{G}^\text{e}_k=\alpha_kG_k^\text{e}\gamma/(KA_0)^2$, $\Bar{\beta}_k\bar{G}_k^\text{e}=\beta_kG_k^\text{e}/(KA_0)$, $\Bar{\alpha}_k\bar{G}_k^\text{id}=\alpha_kG_k^\text{id}/(KA_0)$, and $\Bar{\beta}_k\bar{G}_k^\text{id}=\beta_kG_k^\text{id}/\gamma$ [see Eqs.~(\ref{eq:projected_coeffs}) and (\ref{eq:projected_stress})] for a representative value of $\zeta_V/\gamma=10$. In the solid phase, the rheological response is dominated by the single normal mode $\boldsymbol{\xi}_D$ marked by the arrow in panel (e), which corresponds to the highest value of coefficients $\Bar{\alpha}_k\Bar{G}^\text{e}_k$ and $\Bar{\beta}_k\bar{G}^\text{e}_k$. Note very different values used for the ordinate axes in the top and bottom panels in (e).}
    \label{fig:moduli_honeycomb_interFriction}
\end{figure*}

In the fluid phase (bottom row of Fig.~\ref{fig:moduli_honeycomb}), the spectrum of normal modes can be grouped in a region of zero-modes ($\lambda_k=0$) and two distinct regions with approximately constant values of eigenvalues $\lambda_k$ (two shaded regions in Fig.~\ref{fig:moduli_honeycomb}d). These two regions of non-zero modes set the two characteristic timescales (marked by arrows in Fig.~\ref{fig:moduli_honeycomb}b) of the shear response of hexagonal tilings in the fluid phase, which can be described with the Burgers model. Note that the zero modes do not contribute to the shear response because the value of normalized coefficient $\Bar{\alpha}_k\Bar{G}^\text{e}_k \approx 0$ (see Fig.~\ref{fig:moduli_honeycomb}f). 

We used the normal mode approach to further demonstrate how the ratio between the area and perimeter moduli, $KA_0/\Gamma$, affects the shear rheology of the vertex model. We measured the storage and loss moduli of disordered tilings with different area modulus $K$ as shown in Fig.~\ref{fig:modulus_areaTerm}a,b for two representative values of $p_0$. The simulation results (symbols) show excellent agreement with the predictions of the normal mode analysis (lines). Note that in Fig.~\ref{fig:modulus_areaTerm} we fixed $\Gamma$ and measured the stresses in units~$\Gamma$ and time in units~$\gamma/\Gamma$. We varied the values of $K$ by two orders of magnitude, as indicated by the colorbars for the ratio $KA_0/\Gamma$. Fig.~\ref{fig:modulus_areaTerm}c,d show the corresponding eigenvalues $\lambda_k$ in the ascending order. The eigenvalues $\lambda_k$ for $k>600$ are strongly affected by the magnitude of the area modulus $K$. This indicates that the area term in the vertex model in Eq.~(\ref{eq:vm_energy}) has a stronger contribution to the normal modes of large eigenvalues, which strongly affects the loss moduli at high frequency as a function of $K$ (see Fig.~\ref{fig:modulus_areaTerm}a,b). Fig.~\ref{fig:modulus_areaTerm}e,f shows the corresponding coefficients $\Bar{\alpha}_k\Bar{G}^\text{e}_k$ and $\Bar{\beta}_k\bar{G}^\text{e}_k$ at a representative value of $KA_0/\Gamma=34.641$. In our previous work~\cite{tong2022linear}, we noted that the standard linear solid and Burgers model cannot accurately capture the shear rheological properties for disordered tilings in the vicinity of the solid-fluid transition with $p_0\approx 3.9$. This can be seen in Fig.~\ref{fig:modulus_areaTerm}b for $p_0=3.87$, where the complex rheological behavior is a consequence of the broad spectrum of normal mode eigenvalues $\lambda_k$ (see Fig.~\ref{fig:modulus_areaTerm}d). This reflects one of the key results of the normal mode approach that the full response of the vertex model is the sum of the contributions from all the normal modes, each of which behaves as a standard linear solid. 
 
\subsection{Friction due to the relative motion of neighboring vertices}

\begin{figure*}[t!]
    \centering
    \includegraphics{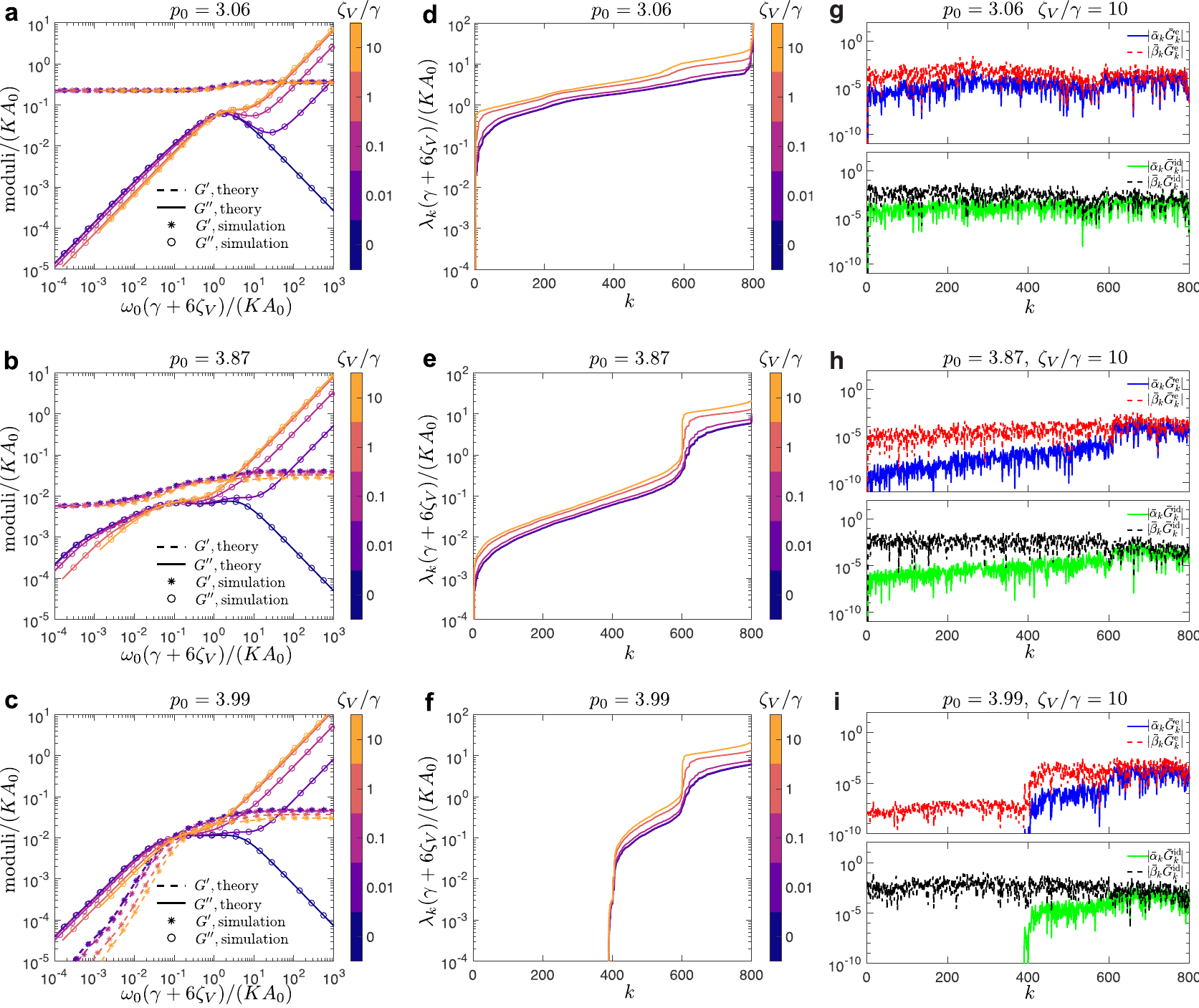}
    \caption{Shear rheology for disordered tilings with internal dissipation due to the relative motion of neighboring vertices in addition to cell-substrate friction. Results for three representative values of the cell shape parameter, $p_0=3.06$ deep in the solid phase (first row), $p_0=3.87$ close to the solid-fluid transition point on the solid side (second row), $p_0=3.99$ in the fluid phase (bottom row) are shown. (a,b,c)~Storage and loss moduli from the simulations~(symbols) compared with the predictions from normal modes~(lines) for different values of the internal friction coefficient $\zeta_V$ (see colorbar). (d,e,f)~Nonzero eigenvalues $\lambda_k$ in ascending order for different values of $\zeta_V$. (g,h,i)~Normalized coefficients $\Bar{\alpha}_k\Bar{G}^\text{e}_k=\alpha_kG_k^\text{e}\gamma/(KA_0)^2$, $\Bar{\beta}_k\bar{G}^\text{e}_k=\beta_kG_k/(KA_0)$, $\Bar{\alpha}_k\bar{G}_k^\text{id}=\alpha_kG_k^\text{id}/(KA_0)$, and $\Bar{\beta}_k\bar{G}_k^\text{id}=\beta_kG_k^\text{id}/\gamma$ for a representative value of $\zeta_V/\gamma=10$.}
    \label{fig:moduli_disordered}
\end{figure*}

In this section, we analyze the shear rheology of the vertex model with the internal friction due to the relative motion of connected vertices characterized by the friction coefficient $\zeta_V$ (see Fig.~\ref{fig:frictionSchematic}b). In addition, the system is subject to vertex-substrate friction with the friction coefficient $\gamma$, discussed in the previous section. In Fig.~\ref{fig:moduli_honeycomb_interFriction}, we first report the results for hexagonal tilings at representative values of the cell shape parameter in the solid phase (top) and the fluid phase (bottom). Fig.~\ref{fig:moduli_honeycomb_interFriction}a,b show an excellent agreement between the storage and loss moduli obtained from simulations~(symbols) and the normal mode analysis~(lines) for various values of the internal friction coefficient $\zeta_V$. Fig.~\ref{fig:moduli_honeycomb_interFriction}e,f show the normalized coefficients $\Bar{\alpha}_k\Bar{G}^\text{e}_k$ and $\Bar{\beta}_k\bar{G}^\text{e}_k$ [see Eq.~(\ref{eq:alpha_k}), Eq.~(\ref{eq:G_e}) and Eq.~(\ref{eq:G_prime})] due to the contribution of elastic forces, $\Bar{\alpha}_k\bar{G}_k^\text{id}=\alpha_kG_k^\text{id}/(KA_0)$ and $\Bar{\beta}_k\bar{G}_k^\text{id}=\beta_kG_k^\text{id}/\gamma$ [see Eq.~(\ref{eq:beta_k}), Eq.~(\ref{eq:G_id}) and Eq.~(\ref{eq:G_double_prime})] due to the contribution of internal dissipation.

The storage and loss moduli in the solid phase are characterized by a single timescale, and all curves for different values of $\zeta_V$ can be collapsed by rescaling the frequency as $\omega_0(\gamma+6\zeta_V)/(KA_0)$. This is because the shear response is dominated by the normal mode $\boldsymbol{\xi}_D$ introduced in the previous section, which is shown by the single peak of the normalized coefficients $\Bar{\alpha}_k\Bar{G}^\text{e}_k$ and $\Bar{\beta}_k\bar{G}^\text{e}_k$ marked by the arrow in Fig.~\ref{fig:moduli_honeycomb_interFriction}e. 
The normal mode $\boldsymbol{\xi}_D$ is simultaneously the eigenvector of the Hessian matrix $\hat{\boldsymbol{H}}$ and the dissipation matrix $\hat{\boldsymbol{C}}$. In particular, $\hat{\boldsymbol{C}}\boldsymbol{\xi}_D=(\gamma+6\zeta_V)\boldsymbol{\xi}_D$, since each vertex in the normal mode $\boldsymbol{\xi}_D$ moves in the opposite direction but with the same magnitude as the three vertices connected to it (see the schematic of $\boldsymbol{\xi}_D$ in the inset of Fig.~\ref{fig:moduli_honeycomb}c). The eigenvalue $\gamma+6\zeta_V$ with respect to $\hat{\boldsymbol{C}}$ thus accounts for the scaling factor of the frequency in Fig.~\ref{fig:moduli_honeycomb_interFriction}a. Note that the values of normalized coefficients $\Bar{\alpha}_k\bar{G}_k^\text{id}$ and $\Bar{\beta}_k\bar{G}_k^\text{id}$ are negligible since the stresses from internal dissipation $\hat{\boldsymbol{\sigma}}^\text{id}_C$ [see Eq.~\ref{eq:stress_id})] cancel out due to hexagonal symmetry. Hence, the rheological behavior in this case is analogous to the one for the hexagonal tiling in the solid phase with only external dissipation  
(compare Fig.~\ref{fig:moduli_honeycomb}a and Fig.~\ref{fig:moduli_honeycomb_interFriction}a).

In the fluid phase, we rescaled the frequency with the same factor to help visually compare the change in the moduli with respect to the friction coefficient $\zeta_V$, as shown in Fig.~\ref{fig:moduli_honeycomb_interFriction}b. Unlike in the solid phase, where the stress contributions $\hat{\boldsymbol{\sigma}}^\text{id}_C$ from internal dissipative forces cancel out due to hexagonal symmetry, there is no such cancellation in the fluid phase and the loss modulus in the fluid phase increases as $G''(\omega)\sim\zeta_V \omega$ at high frequencies $\omega$ [see Eq.~(\ref{eq:G_double_prime})]. This is because the hexagonal state is unstable in the fluid regime and cells become slightly distorted in a local energy minimum ~\cite{tong2022linear}. Note that the crossover to the asymptotic regime $G''(\omega)\sim\zeta_V \omega$   shifts to lower frequencies as the ratio $\zeta_V/\gamma$ increases. 
 
 \begin{figure*}[t!]
    \centering
    \includegraphics{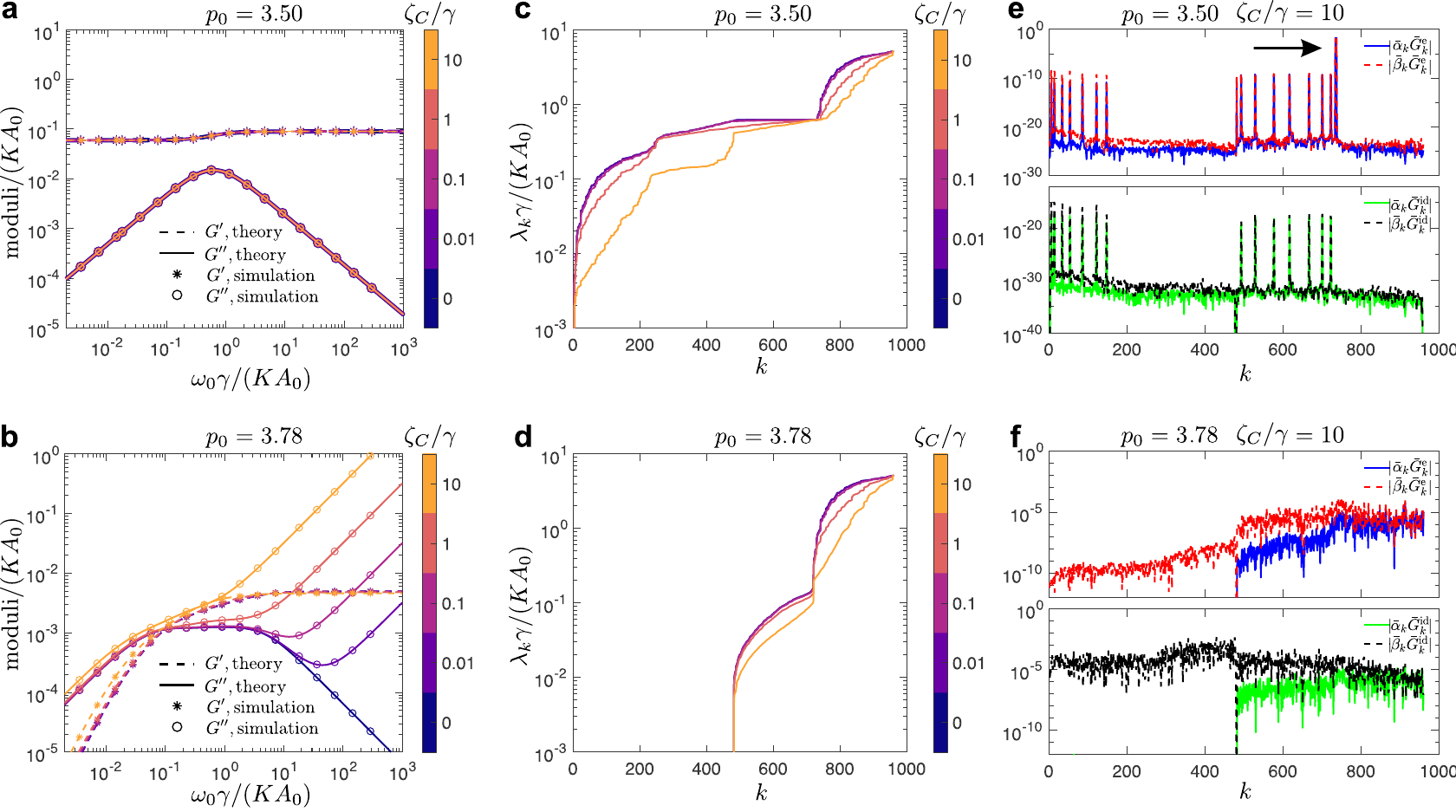}
    \caption{Shear rheology of hexagonal tilings with internal dissipation due to relative motion of neighboring cell centers and external cell-substrate friction. Results for two representative values of the cell shape parameter, $p_0=3.5$ in the solid phase (top row) and $p_0=3.78$ in the fluid phase (bottom row) are shown. (a,b)~Storage and loss moduli from the simulations (symbols) compared with the predictions of the normal mode analysis (lines) for different values of $\zeta_C/\gamma$ (see colorbar). (c,d)~Nonzero eigenvalues $\lambda_k$ in ascending order for different values of $\zeta_C/\gamma$. (e,f)~Normalized coefficients $\Bar{\alpha}_k\Bar{G}^\text{e}_k=\alpha_kG_k^\text{e}\gamma/(KA_0)^2$, $\Bar{\beta}_k\bar{G}^\text{e}_k=\beta_kG_k/(KA_0)$, $\Bar{\alpha}_k\bar{G}_k^\text{id}=\alpha_kG_k^\text{id}/(KA_0)$, and $\Bar{\beta}_k\bar{G}_k^\text{id}=\beta_kG_k^\text{id}/\gamma$ for a representative value of $\zeta_C/\gamma=10$. In the solid phase, the rheological response is dominated by the single normal mode $\boldsymbol{\xi}_D$ marked by the arrow in panel (e). This mode corresponds to the highest value of coefficients $\Bar{\alpha}_k\Bar{G}^\text{e}_k$ and $\Bar{\beta}_k\bar{G}^\text{e}_k$. Note very different numerical values on ordinate axes in top and bottom panels in~(e). }
    \label{fig:moduli_honeycomb_cell}
\end{figure*}

We proceed to perform simulations of the vertex model with disordered tilings, which mimic the geometry of real epithelial tissues.
The first column in Fig.~\ref{fig:moduli_disordered} shows the storage and loss moduli from simulations~(symbols) for representative values of the cell shape parameter deep in the solid phase (top row, $p_0=3.06$), close to the solid-fluid transition point on the solid side (middle row, $p_0=3.87$), and in the fluid phase (bottom row, $p_0=3.99$), which show excellent agreement with predictions of the normal mode analysis~(lines). As usual, the frequency is again measured in units of $KA_0/(\gamma+6\zeta_V)$. When the system is deep in the solid phase, as shown in Fig.~\ref{fig:moduli_disordered}a, the response is similar to that of a standard linear solid, with loss modulus having one peak for small values of $\zeta_V/\gamma$ in the low-frequency regime, and crossing over to the asymptotic behavior $G''(\omega)\sim\zeta_V \omega$ in the high-frequency regime, where the internal dissipation dominates. Increasing the ratio $\zeta_V/\gamma$ moves this crossover to lower frequencies, which is analogous to the behavior of hexagonal tilings in the fluid phase (see Fig.~\ref{fig:moduli_honeycomb_interFriction}b). 

When the system is close to the solid-fluid transition, but on the solid side (see Fig.~\ref{fig:moduli_disordered}b), the loss modulus develops two peaks, whose separation becomes more pronounced as $\zeta_V/\gamma$ increases. This is similar to the behavior for hexagonal tilings shown in Fig.~\ref{fig:moduli_honeycomb_interFriction}b, which can be accounted for by the jump between the two regions of eigenvalues $\lambda_k$ becoming sharper with increasing $\zeta_V/\gamma$, as shown in Figs.~\ref{fig:moduli_honeycomb_interFriction}d and \ref{fig:moduli_disordered}e. The internal dissipation, however, dominates in the high-frequency regime, and the loss modulus crosses over to increasing with frequency, which is again shifted to the left as $\zeta_V/\gamma$ increases. Furthermore, the loss modulus crosses over from the linear scaling ($\sim \omega$) at low frequencies to anomalous scaling ($\sim \omega^\alpha$) with the fractional exponent $\alpha<1$ at intermediate frequencies (see Fig.~\ref{fig:moduli_disordered}b), which was already noted in our prior work in Ref.~\cite{tong2022linear}.
This is because the eigenvalues $\lambda_k$ gradually increase by two orders of magnitude up to the sharp jump point in Fig.~\ref{fig:moduli_disordered}e.

 \begin{figure*}
    \centering
    \includegraphics{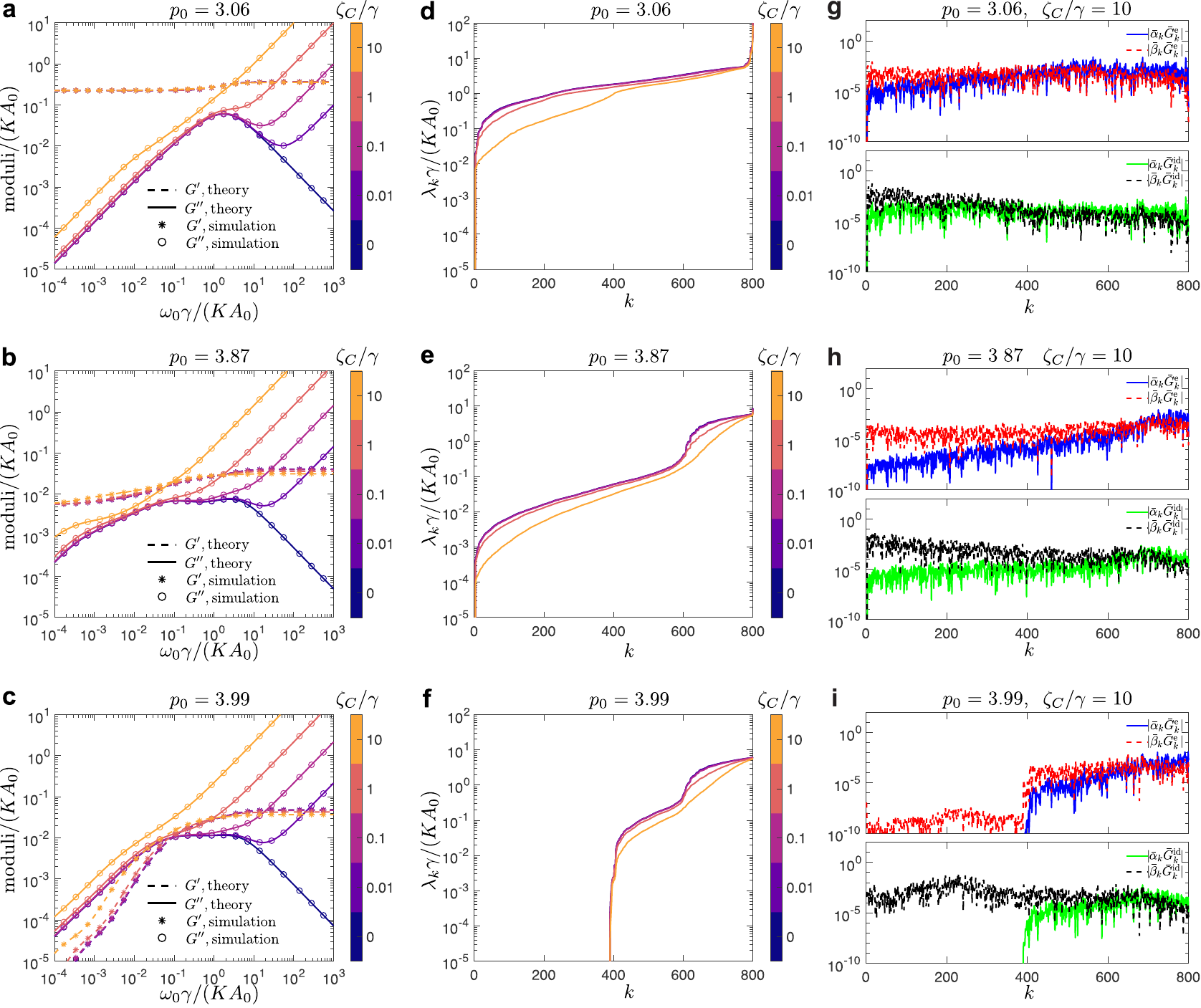}
    \caption{Shear rheology for disordered tilings with internal dissipation due to the relative motion of neighboring cell centers and external  cell-substrate friction. Results are shown for three representative values of the cell shape parameter, $p_0=3.06$ deep in the solid phase (top row), $p_0=3.87$ close to the solid-fluid transition point on the solid side (middle row), and $p_0=3.99$ in the fluid phase (bottom row). (a,b,c)~Storage and loss moduli from the simulations~(symbols) compared with the predictions of the normal mode analysis~(lines) for different values of $\zeta_C/\gamma$ (see colorbar). (d,e,f)~Nonzero eigenvalues $\lambda_k$ in ascending order for different values of $\zeta_C/\gamma$. (g,h,i)~Normalized coefficients $\Bar{\alpha}_k\Bar{G}^\text{e}_k=\alpha_kG_k^\text{e}\gamma/(KA_0)^2$, $\Bar{\beta}_k\bar{G}^\text{e}_k=\beta_kG_K/(KA_0)$, $\Bar{\alpha}_k\bar{G}_k^\text{id}=\alpha_kG_k^\text{id}/(KA_0)$, and $\Bar{\beta}_k\bar{G}_k^\text{id}=\beta_kG_k^\text{id}/\gamma$ at one representative value of $\zeta_C/\gamma=10$.}
    \label{fig:moduli_disordered_cell}
\end{figure*}

In the fluid phase, the internal dissipation $\zeta_V$ has a similar effect on the loss modulus as shown in Fig.~\ref{fig:moduli_disordered}c and one observes that the loss modulus crosses over to growing linearly with frequency in the high-frequency regime. 

\subsection{Friction due to relative motions of neighboring cells}

Finally, we investigated the shear rheology of the vertex model with the internal friction due to the relative motion of neighboring cell centers characterized by the friction coefficient $\zeta_C$ (see Fig.~\ref{fig:frictionSchematic}c). As in the previous section, vertices also experience friction with the substrate with the friction coefficient $\gamma$. In Fig.~\ref{fig:moduli_honeycomb_cell}, we show the dynamic moduli for hexagonal tilings at representative values of the cell shape parameter in the solid (top) and the fluid phase (bottom). In the solid phase, the dominant normal mode $\boldsymbol{\xi}_D$ does not generate friction due to relative motion of neighboring cells, i.e., $\hat{\boldsymbol{C}}\boldsymbol{\xi}_D=\gamma\boldsymbol{\xi}_D$. Thus, the shear rheology for the hexagonal tiling in the solid phase is identical to that of the model with the cell-substrate friction only, cf.~Fig.~\ref{fig:moduli_honeycomb}. 

In the fluid phase, the two small humps in the loss modulus that are present in the case with only the vertex-substrate dissipation (i.e., for $\zeta_C=0$) are smoothed out as $\zeta_C/\gamma$ increases (see Fig.~\ref{fig:moduli_honeycomb_cell}b). This is because the jump between two regions of eigenvalues $\lambda_k$ are smoothed out with increasing $\zeta_C/\gamma$, as shown in Fig.~\ref{fig:moduli_honeycomb_cell}d. Note that this is different from the  the case with dissipation due to the relative motion of neighboring vertices, where the jump in the spectrum of eigenvalues $\lambda_k$ becomes more pronounced as the value of $\zeta_V/\gamma$ increases (see Fig.~\ref{fig:moduli_honeycomb_interFriction}d) 

In Fig.~\ref{fig:moduli_disordered_cell}, we report the results for disordered tilings. The first column in Fig.~\ref{fig:moduli_disordered_cell} shows the storage and loss moduli from simulations~(symbols) for representative values of the cell shape parameter deep in the solid phase (top row, $p_0=3.06$), close to the solid-fluid transition point on the solid side (middle row, $p_0=3.87$), and in the fluid phase (bottom row, $p_0=3.99$), which all show excellent match with predictions from the normal modes~(lines). For a system deep in the solid phase, the loss modulus develops a secondary shoulder at low frequencies as $\zeta_C$ increases (see Fig.~\ref{fig:moduli_disordered_cell}a), which corresponds to the jump of the eigenvalues $\lambda_k$ developing at $k\approx400$ in Fig.~\ref{fig:moduli_disordered_cell}d. 

When the system is close to the solid-fluid transition on the solid side, the range of intermediate frequencies over which the loss modulus has a significant value becomes wider with a small bump developing at low frequency as $\zeta_C/\gamma$ increases (see Fig.~\ref{fig:moduli_disordered_cell}b). This is reflected by a wider range of eigenvalues $\lambda_k$ with increasing $\zeta_C/\gamma$ in Fig.~\ref{fig:moduli_disordered_cell}e. This behavior is different from the response of the system with friction due to the relative motion of neighboring vertices, for which the dynamic range of characteristic time scales does not significantly change with $\zeta_V$ (see Fig.~\ref{fig:moduli_disordered}e).

In the fluid phase, similarly to the hexagonal tilings, the two small bumps in the loss modulus at intermediate frequencies, when there is only external dissipation (i.e., for $\zeta_C=0$), are smoothed out as $\zeta_C/\gamma$ increases (see Fig.~\ref{fig:moduli_disordered_cell}c), which is reflected by the jump of eigenvalues $\lambda_k$ at $k\approx600$ being smoothed out with increasing $\zeta_C$ as shown in Fig.~\ref{fig:moduli_disordered_cell}f.

\section{Discussion and Conclusions}
\label{sec:discussion}
In this work, we used the normal modes formalism to develop a general method for calculating dynamic rheological moduli of soft materials in the linear response regime in the presence of internal and external dissipation. A key finding is that the dynamic rheological moduli can be represented as a linear superposition of standard linear solid and Jeffreys elements connected in parallel, with each mode contributing a characteristic relaxation timescale related to its eigenvalue. The external and internal dissipation, i.e., friction with the substrate and friction between constitutive elements of the system, respectively, have markedly different effects on the rheology. The external dissipation together with elastic relaxation combine to result in the standard linear solid model, i.e., a viscoelastic solid. On the other hand, the internal dissipation is described by the Jeffreys element, and it represents a viscoelastic liquid. As the result, the behavior of the loss modulus is qualitatively different at high frequencies depending on whether the internal dissipation is present or not.  While the calculated dynamical moduli depend on the precise details of microscopic dynamics and dissipation mechanism, the method presented here based on normal modes  is agnostic to such details. It can be applied to systems with any number of spatial dimensions as long as it is possible to define an energy function with well-defined local minima that is differentiable twice. Although here we considered the overdamped case, extension to systems with inertia is straightforward. Finally, for systems with less than $\sim10^4$ degrees of freedom, this method is superior in terms of computational cost compared to direct simulations since one needs to solve the generalized eigenvalue problem only once, which can then be used to determine the linear response properties over the full range of frequencies. 

We applied this formalism to study the linear response to shear deformations of the two-dimensional vertex model for epithelial tissue mechanics with three different microscopic mechanisms of dissipation. We derived expressions for mechanical stresses on cells due to elastic and dissipative forces and showed that for all three dissipation models, the method gives an excellent agreement with direct numerical simulations.  Although our analysis of the vertex model is limited to the linear response regime and is unable to capture the response to large deformations, especially if those involve local plastic rearrangements, it nonetheless provides valuable insights into its complex rheology. In particular, it allows one to fully understand the behavior of the storage and loss moduli of the vertex model in the linear response regime in terms of the behaviors of each normal mode. Applying this approach to compute and understand other response functions, e.g., the bulk modulus is also straightforward, as is the treatment of different forms of external driving. For example, modeling a typical experimental setup where the system is clamped at two of its ends that are then moved relative to each other would just involve introducing the appropriate functional form for the driving force in Eq.~(\ref{eq:motion_linear}). Additionally, the generalized eigenvalue problem in Eq.~(\ref{eq:eigen_problem}) would have to be solved subject to appropriate boundary conditions. 

The ability to study the effects of both internal and external dissipation at the same footing makes this approach appealing to studying the effects of complex dissipative processes. In addition, the method can be directly extended to models that include the effects of activity.~\cite{staddon2019mechanosensitive, tlili2019shaping, sknepnek2021generating, comelles2021epithelial, krajnc2021active} Understanding the roles of activity and internal dissipation is crucial for a proper understanding of the rheology of living tissues.

\begin{acknowledgments}
This research was primarily supported by NSF through Princeton University’s Materials Research Science and Engineering Center DMR-2011750 and by the Project X Innovation Research Grant from the Princeton School of Engineering and Applied Science (S.T. and A.K.). R.S. acknowledges support by the UK BBSRC (Award BB/N009789/1) and the UK EPSRC (Award EP/W023946/1). This collaboration was initiated during the KITP program ``Symmetry, Thermodynamics and Topology in Active Matter'' (ACTIVE20), and it is supported in part by the National Science Foundation under Grant No.\ NSF PHY-1748958.
\end{acknowledgments}

\appendix 


\section{Elastic force on a vertex}
\label{app:force-on-vertex}
In this appendix, we outline the derivation of the expression for the elastic force $\mathbf{f}_i^\text{e}$ on a vertex $i$. 
The elastic force on vertex $i$ is 
\begin{equation}
  \mathbf{f}_i^\text{e}=-\nabla_{\mathbf{R}_i}E,  
\end{equation}
where
\begin{equation}   E=\sum_{C}\left[\frac{K}{2}\left(A_C-A_{0}\right)^2+\frac{\Gamma}{2}\left(P_C-P_{0}\right)^2\right]
\end{equation}
is the energy function of the vertex model. In these appendices, we use Latin subscript indices to denote different vertices and cells, and Greek superscript indices to denote $x$ and $y$ components of a vector.

The area of the cell $C$ can be expressed as
\begin{eqnarray}
A_{C}&=&\frac{1}{2}\sum_{i\in C}\left(\mathbf{R}_{i}\times\mathbf{R}_{i+1}\right)\cdot \mathbf{e}_{z},\nonumber \\
A_{C}&=&\frac{1}{2}\sum_{i\in C}\varepsilon_{\alpha\beta}R^{\alpha}_{i}R^{\beta}_{i+1}.
\label{eq:app:area}
\end{eqnarray}
In the above equations, the summation is over all vertices $i$ that belong to cell $C$. Vertices $i$ are assumed to be labeled from $0$ to $N_C-1$ in a counterclockwise sense and the index $i+1$ is calculated modulo $N_C$, i.e., for $i=N_C-1$ the value of $i+1$ is $0$. In Eq.~\eqref{eq:app:area}, the $\mathbf{R}_i$ is the position vector of vertex $i$, $\mathbf{e}_{z}$ is the unit-length vector perpendicular to the plane of the tissue (assumed to be the $xy-$plane),  $\varepsilon_{\alpha\beta}$ is the two-dimensional Levi-Civita symbol, and summation over repeated Greek indices is implied. Similarly, the perimeter of the cell $C$ is
\begin{eqnarray}
    P_{C}&=&\sum_{i\in C}\left|\mathbf{R}_{i+1}-\mathbf{R}_{i}\right| \nonumber \\
         &=&\sum_{i\in C}\big[\left(R^{\alpha}_{i+1}-R^{\alpha}_{i}\right)\left(R^{\alpha}_{i+1}-R^{\alpha}_{i}\right)\big]^{1/2},\label{eq:perimeter}
\end{eqnarray}
with the same rules as in the case of the area term in Eq.~(\ref{eq:app:area}).

The elastic force on vertex $i$ is then
\begin{eqnarray}
    \mathbf{f}^{\mathrm{e}}_i &=& -\sum_{C \in \mathcal{N}_i}\big[ K\left(A_{C}-A_{0}\right)\left(\nabla_{\mathbf{R}_{i}}A_{C}\right) \nonumber \\
    && \quad \quad \quad \quad +\, \Gamma\left(P_{C}-P_{0}\right)\left(\nabla_{\mathbf{R}_{i}}P_{C}\right)\big], \label{eq:force_vert}\\
\mathbf{f}^{\mathrm{e}}_i &\equiv& \sum_{C \in \mathcal{N}_i} \mathbf{f}^{\mathrm{e}}_{C\rightarrow i} \nonumber
\end{eqnarray}
where $\mathcal{N}_i$ includes the set of all cells that share vertex $i$ and we defined $\mathbf{f}^{\mathrm{e}}_{C\rightarrow i}$ as the elastic force contribution on the vertex $i$ due to cell $C$.

It is straightforward to show that for the vertex $i$ that belongs to cell $C$ the derivatives are 
\begin{eqnarray}
        \nabla_{\mathbf{R}_{i}}A_{C} &=&  \frac{1}{2} \delta_{C,i} \left(\mathbf{R}_{i+1}-\mathbf{R}_{i-1}\right)\times\mathbf{e}_{z} ,\nonumber \\ 
        \frac{\partial A_C}{\partial R_i^\alpha} & = &  \frac{1}{2} \delta_{C,i} \varepsilon_{\alpha\beta} \left(R_{i+1}^\beta - R_{i-1}^\beta \right), \label{eq:grad_area}  \\ 
         \frac{\partial A_C}{\partial R_i^\alpha} & = &  \frac{1}{2} \delta_{C,i} \varepsilon_{\alpha\beta} \left(l_{i,i+1}^\beta - l_{i,i-1}^\beta \right), \nonumber
\end{eqnarray}
and
\begin{eqnarray}
    \nabla_{\mathbf{R}_{i}}P_{C} &=& \delta_{C,i} \left( \frac{\mathbf{R}_{i} - \mathbf{R}_{i-1}}{|\mathbf{R}_{i} - \mathbf{R}_{i-1}|} - \frac{\mathbf{R}_{i+1} - \mathbf{R}_{i}}{|\mathbf{R}_{i+1} - \mathbf{R}_{i}|} \right), \nonumber\\ 
    \frac{\partial P_C}{\partial R_i^\alpha} & = & \delta_{C,i} \left( \frac{R^\alpha_{i} - R^\alpha_{i-1}}{|\mathbf{R}_{i} - \mathbf{R}_{i-1}|} - \frac{R^\alpha_{i+1} - R^\alpha_{i}}{|\mathbf{R}_{i+1} - \mathbf{R}_{i}|} \right), 
    \label{eq:grad_perim} \\
    \frac{\partial P_C}{\partial R_i^\alpha} & = & -\delta_{C,i} \left( \hat{l}^\alpha_{i,i-1}+ \hat{l}^\alpha_{i,i+1}\right), \nonumber
\end{eqnarray}
where we introduced the vector $\mathbf{l}_{i,j}=\mathbf{R}_{j}-\mathbf{R}_{i}$ along the junction connecting vertices $i$ and $j$, normalized unit vector $\hat{\mathbf{l}}=\mathbf{l}/\left|\mathbf{l}\right|$, 
and
\begin{equation}
    \delta_{C,i}=\begin{cases}
                        1 \;\;\;\text{if vertex}\;i\;\text{belongs to cell}\;C \\
                        0 \;\;\;\text{otherwise}
                  \end{cases}.   
\end{equation}
The elastic force contribution $\mathbf{f}^{\mathrm{e}}_{C\rightarrow i}$ on the vertex $i$ due to cell $C$ can thus be expressed as  
\begin{eqnarray}
\mathbf{f}^{\mathrm{e}}_{C\rightarrow i}&=&-\frac{1}{2} \delta_{C,i}\, K\left(A_{C}-A_{0}\right) \left(\mathbf{l}_{i,i+1}-\mathbf{l}_{i,i-1}\right)\times\mathbf{e}_{z} \nonumber \\
&& + \delta_{C,i}\, \Gamma\left(P_{C}-P_{0}\right) \left(\mathbf{\hat{l}}_{i,i-1}+\hat{\mathbf{l}}_{i,i+1}\right) \label{eq:elastic_force_from_cel_to_vertex}
\end{eqnarray}

\begin{figure}
    \centering
    \includegraphics[width=0.95\columnwidth]{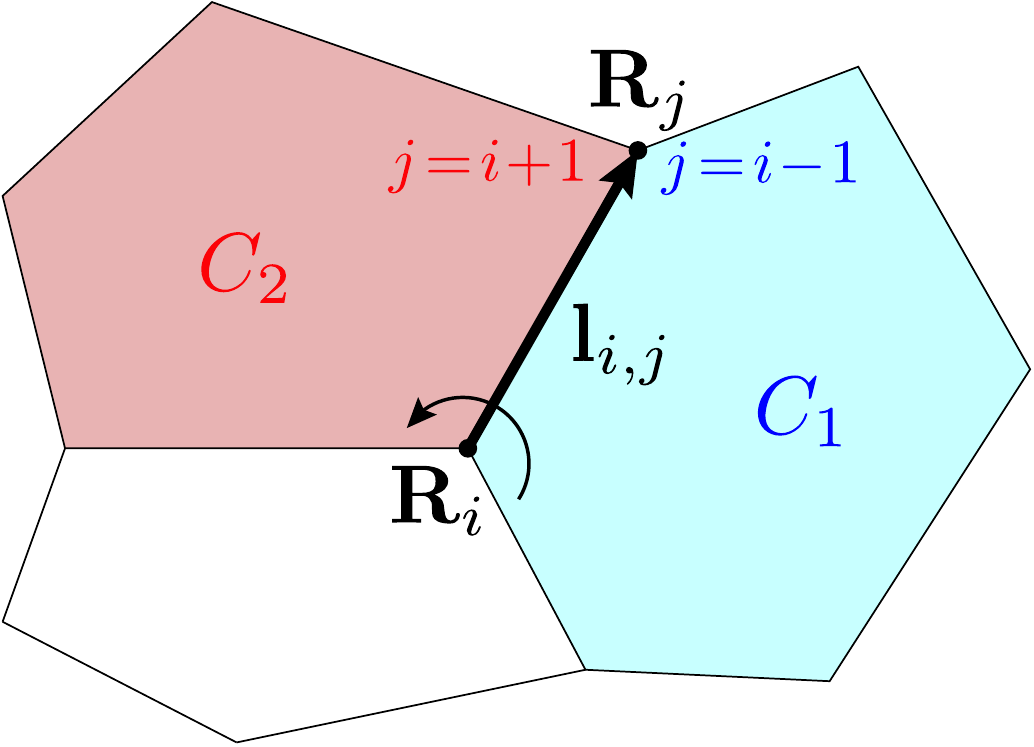}
    \caption{An efficient way of calculating the elastic force on vertex $i$, $\mathbf{f}_i^\text{e}$, is to loop over all cell-cell junctions that originate at $i$ in the counterclockwise direction. For each junction, the two cells that share it contribute to the total force. These contributions are the two terms in the sum in Eq.~(\ref{eq:force_on_vertex_i}). For consistency, we adopt a convention that when looking along the junction away from the vertex $i$, the cell to the right (blue) is labeled as $C_1$, and the cell to the left (red) is labeled as $C_2$. Note that since vertices within each cell are ordered counterclockwise, the endpoint of the junction, i.e., the vertex $j$ appears in cell 1 (2) as $i-1$ ($i+1$) in the cell's internal labeling.  }
    \label{fig:cell_loop}
\end{figure}
Note that the derivatives $\nabla_{\mathbf{R}_{i}}A_{C}$ and $\nabla_{\mathbf{R}_{i}}P_{C}$ only depend on the lengths and orientations of cell-cell junctions that contain vertex $i$
[see Eqs.~(\ref{eq:grad_area}) and (\ref{eq:grad_perim})]. Thus the elastic force $\mathbf{f}^{\mathrm{e}}_i$ can also be expressed as a summation over the cell-cell junctions as
\begin{eqnarray}
    \mathbf{f}^{\mathrm{e}}_{i}=&+&\sum_{j\in\mathcal{S}_i}\frac{1}{2} K \left(A_{C_1(j)}-A_{C_2(j)}\right)\mathbf{l}_{i,j} \times \mathbf{e}_{z}\nonumber\\
    &+&\sum_{j\in\mathcal{S}_i}\Gamma\left(\Delta P_{C_1(j)} + \Delta P_{C_2(j)}\right)\hat{\mathbf{l}}_{i,j},\label{eq:force_on_vertex_i}
\end{eqnarray}
where $\Delta P_{C}=\left(P_{C}-P_{0}\right)$. The set $\mathcal{S}_i$ contains all vertices, $j$, connected by cell-cell junctions to the vertex $i$. Here, we adopt a convention that when looking along the junction away from the vertex $i$, the cells to the right and left are labeled as $C_1(j)$ and $C_2(j)$, respectively (see Fig.~\ref{fig:cell_loop}). In the numerical implementation, we calculated the elastic force  $\mathbf{f}^{\mathrm{e}}_{i}$ using Eq.~(\ref{eq:force_on_vertex_i}) because we can efficiently loop over all cell-cell junctions that belong to the vertex $i$.


\section{Hessian matrix of the vertex model}\label{appendix:Hessian}
In this appendix, we outline the derivation of the expressions for the Hessian matrix of the vertex model. To allow for the most general case, we express the energy of the vertex model as
\begin{equation}
    E=\sum_{C}\frac{K_{C}}{2}\left(A_{C}-A_{0}^{C}\right)^{2}+\sum_{C}\frac{\Gamma_{C}}{2}P_{C}^{2}-\sum_{\left\langle i,j\right\rangle }\Lambda_{ij} |\mathbf{l}_{i,j}|,\label{eq:VM-full}
\end{equation}
where $\Lambda_{ij}$ is the line tension and $|\mathbf{l}_{i,j}|\equiv |\mathbf{R}_{j}-\mathbf{R}_{i}|$ is the length of the junction connecting vertices $i$ and $j$. The summation $\langle i,j \rangle$ is over all pairs of vertices $i$ and $j$ connected by junctions. In the case when parameters $K_C$, $\Gamma_C$, $A_0^C$, $P_0^C$ and $\Lambda_{ij}$ are not cell-specific, one immediately reads off $P_0=\frac{\Lambda}{\Gamma}$ and, after adding the constant term $\frac{1}{2}\Gamma P_0^2$, readily recovers Eq.~(\ref{eq:vm_energy}). Since in general scenarios the line tension can vary across junctions, we will use Eq.~(\ref{eq:VM-full}) as the expression for the energy of the vertex model for further derivation.

For a system with $N$ vertices, the Hessian of the vertex model is a real symmetric matrix of size $2N\times2N$. Its elements are
\begin{equation}
    \hat{H}_{IJ}=\frac{\partial^{2}E}{\partial x_{I}\partial x_{J}}\Big|_{\mathbf{r}=\mathbf{r}^{\text{eq}}}, \label{eq:app:Hessian}
\end{equation}
where $I,J\in\{1,\dots,2N\}$, and $I=2(i-1)+\alpha$, with $i\in\{1,\dots,N\}$ and $\alpha\in\{1,2\}$. In other words, for $I=2i-1$, $x_I\equiv R_i^x$ and for $I=2i$, $x_I\equiv R_i^y$. Identical relations hold for index $J$. Elements of the Hessian matrix are calculated for a configuration in mechanical equilibrium, i.e., for  ${\mathbf{r}=\mathbf{r}^{\text{eq}}}$. One finds,
\begin{eqnarray}
    \frac{\partial^{2}E}{\partial R_{k}^{\alpha}\partial R_{m}^{\beta}}&=&\sum_{C} K_{C} \left[\frac{\partial A_{C}}{\partial R_{k}^{\alpha}}\frac{\partial A_{C}}{\partial R_{m}^{\beta}}
    +\left(A_{C}-A_{0}^{C}\right)\frac{\partial^{2}A_{C}}{\partial R_{k}^{\alpha}\partial R_{m}^{\beta}}\right]\nonumber\\&&+\sum_{C}\Gamma_{C}\left[\frac{\partial P_{C}}{\partial R_{k}^{\alpha}}\frac{\partial P_{C}}{\partial R_{m}^{\beta}}
    +\sum_{C}\Gamma_{C}P_{C}\frac{\partial^{2}P_{C}}{\partial R_{k}^{\alpha}\partial R_{m}^{\beta}}\right]\nonumber\\
    &&-\sum_{\left\langle i,j\right\rangle }\Lambda_{ij}\frac{\partial^{2} |\mathbf{l}_{i,j}|}{\partial R_{k}^{\alpha}\partial R_{m}^{\beta}}.
\end{eqnarray}

The expression for $\partial A_{C}/\partial R_{k}^{\alpha}$ and $\partial P_{C}/\partial R_{k}^{\alpha}$ are given in Eqs.~(\ref{eq:grad_area}) and (\ref{eq:grad_perim}), respectively. It is straightforward to show that
\begin{equation}
    \frac{\partial^{2}A_{C}}{\partial R_{k}^{\alpha}\partial R_{m}^{\beta}}=\frac{1}{2} \delta_{C,k} \delta_{C,m}\varepsilon_{\alpha\beta} \left(\delta_{m-1,k}-\delta_{m+1,k}\right),
\end{equation}
where 
\begin{equation}
 \delta_{i,j} = \begin{cases}
                        1 \;\;\text{if}\;\; i=j \\
                        0 \;\;\text{if}\;\; i\neq j
                    \end{cases}    
\end{equation}
is the Kronecker delta. 
\begin{widetext}
Similarly,
\begin{eqnarray}
    \frac{\partial^2 P_{C}}{\partial R_{k}^{\alpha}\partial R_{m}^{\beta}}&=&-\delta_{C,k} \delta_{C,m} \left(\delta_{m,k}-\delta_{m-1,k}\right) \left[\frac{ \left(R_{m}^{\alpha}-R_{m-1}^{\alpha}\right)\left(R_{m}^{\beta}-R_{m-1}^{\beta}\right)}{\left|\mathbf{R}_{m}-\mathbf{R}_{m-1}\right|^{3}}
    -\frac{\delta_{\alpha\beta}}{\left|\mathbf{R}_{m}-\mathbf{R}_{m-1}\right|} \right]\nonumber\\
    &&+\delta_{C,k} \delta_{C,m}  \left(\delta_{m+1,k}-\delta_{m,k}\right) \left[\frac{\left(R_{m+1}^{\alpha}-R_{m}^{\alpha}\right)\left(R_{m+1}^{\beta}-R_{m}^{\beta}\right)}{\left|\mathbf{R}_{m+1}-\mathbf{R}_{m}\right|^{3}}
    -\frac{\delta_{\alpha\beta}}{\left|\mathbf{R}_{m+1}-\mathbf{R}_{m}\right|}\right], \\
        \frac{\partial^2 P_{C}}{\partial R_{k}^{\alpha}\partial R_{m}^{\beta}}&=&-\delta_{C,k} \delta_{C,m} \left(\delta_{m,k}-\delta_{m-1,k}\right) \left[\frac{ l_{m-1,m}^{\alpha}l_{m-1,m}^{\beta}}{\left|\mathbf{l}_{m-1,m}\right|^{3}}
    -\frac{\delta_{\alpha\beta}}{\left|\mathbf{l}_{m-1,m}\right|} \right]\nonumber\\
    &&+\delta_{C,k} \delta_{C,m}  \left(\delta_{m+1,k}-\delta_{m,k}\right) \left[\frac{l_{m,m+1}^{\alpha}l_{m,m+1}^{\beta}}{\left|\mathbf{l}_{m,m+1}\right|^{3}}
    -\frac{\delta_{\alpha\beta}}{\left|\mathbf{l}_{m,m+1}\right|}\right].\nonumber
\end{eqnarray}
Finally,
\begin{eqnarray}
    \frac{\partial^{2}|\mathbf{l}_{i,j}|}{\partial R_{k}^{\alpha}\partial R_{m}^{\beta}}&=&- \left(\delta_{i,k}-\delta_{j,k}\right) \left(\delta_{i,m}-\delta_{j,m}\right)\left[\frac{\left(R_{j}^{\alpha}-R_{i}^{\alpha}\right) \left(R_{j}^{\beta}-R_{i}^{\beta}\right)}{\left|\mathbf{R}_{j}-\mathbf{R}_{i}\right|^{3}}-\frac{\delta_{\alpha\beta}}{\left|\mathbf{R}_{j}-\mathbf{R}_{i}\right|}\right], \\
    \frac{\partial^{2}|\mathbf{l}_{i,j}|}{\partial R_{k}^{\alpha}\partial R_{m}^{\beta}}&=&- \left(\delta_{i,k}-\delta_{j,k}\right) \left(\delta_{i,m}-\delta_{j,m}\right)\left[\frac{l_{i,j}^{\alpha} l_{i,j}^{\beta}}{\left|\mathbf{l}_{i,j}\right|^{3}}-\frac{\delta_{\alpha\beta}}{\left|\mathbf{l}_{i,j}\right|}\right].\nonumber
\end{eqnarray}
\end{widetext}


\section{External driving force due to shear of the periodic simulation box in the vertex model}\label{appendix:driving_fb}
In this appendix, we outline the derivation of the expression for the external driving force $\bar{\mathbf{f}}^{\text{pb}}$ due to the shear of the rectangular periodic simulation box with edge lengths  $\ell_x$ and $\ell_y$. We use the energy function for the vertex model given in Eq.~(\ref{eq:VM-full}) and seek to find the expression for 
\begin{equation}
    \bar{f}^\text{pb}_{I}=-\frac{\partial^{2}E}{\partial x_{I}\partial \epsilon}\Big|_{\mathbf{r}=\mathbf{r}^{\text{eq}},\epsilon=0},
\end{equation}
where $\epsilon$ measures the shear of the simulation box, and as in Eq.~(\ref{eq:app:Hessian}), $I\in\{1,\dots,2N\}$ labels vertex coordinates.
Similar to the derivation of the Hessian matrix given in Appendix \ref{appendix:Hessian}, one finds
\begin{eqnarray}
\frac{\partial^{2}E}{\partial R_{k}^{\alpha}\partial\epsilon} &=&\sum_{C}K_{C}\left[\frac{\partial A_{C}}{\partial R_{k}^{\alpha}}\frac{\partial A_{C}}{\partial \epsilon}
    +\left(A_{C}-A_{0}^{C}\right)\frac{\partial^{2}A_{C}}{\partial R_{k}^{\alpha}\partial \epsilon}\right]\nonumber \\
    &&
    +\sum_{C}\Gamma_{C}\left[\frac{\partial P_{C}}{\partial R_{k}^{\alpha}}\frac{\partial P_{C}}{\partial \epsilon}
    +P_{C}\frac{\partial^{2}P_{C}}{\partial R_{k}^{\alpha}\partial \epsilon}\right]\nonumber\\
    &&-\sum_{\left\langle i,j\right\rangle }\Lambda_{ij}\frac{\partial^{2}|\mathbf{l}_{i,j}|}{\partial R_{k}^{\alpha}\partial \epsilon},
\end{eqnarray}
with the same meaning of the summation indices. 
Note that the shear degree of freedom $\epsilon$ appears only for cell junctions that cross the periodic boundary. The energy of the vertex model depends on $\epsilon$ through the $x$-component of the distance vectors, 
\begin{equation}
    l_{m,n}^x = R_{n}^x-R_{m}^x+q_{m,n}^{x}\ell_{x}+\epsilon q_{m,n}^{y}\ell_{y},
\end{equation}
where~\cite{merkel2018geometrically}
\begin{subequations}
\begin{align}
            q_{m,n}^{x} &= \begin{cases}
                       \hphantom{+1\,}\mathrlap0\hphantom{(-1)} \;\;\parbox{0.28\textwidth}{if the junction connecting vertices $m$ and $n$ does not cross the right or left boundaries} \vspace{0.1cm}\\[3ex]
                       +1\,(-1) \;\;\parbox{0.28\textwidth}{if the junction connecting vertices $m$ and $n$ crosses the right (left) boundary} 
                    \end{cases},\\
    q_{m,n}^{y} &= \begin{cases}
                       \hphantom{+1\,}\mathrlap0\hphantom{(-1)} \;\;\parbox{0.28\textwidth}{if the junction connecting vertices $m$ and $n$ does not cross the top or bottom boundaries} \\[3ex]
                       +1\, (-1) \;\;\parbox{0.28\textwidth}{if the junction connecting vertices $m$  and crosses the top (bottom) boundary} 
                    \end{cases}.
\end{align}
\end{subequations}
Using the chain rule, the derivative with respect to $\epsilon$ is 
\begin{equation}
    \frac{\partial}{\partial\epsilon}=\sum_{\langle m,n\rangle}\frac{\partial l_{m,n}^x}{\partial\epsilon}\frac{\partial}{\partial l_{m,n}^x}=\sum_{\langle m,n\rangle}q_{m,n}^{y}\ell_{y}\frac{\partial}{\partial l_{m,n}^x}.
    \label{eq:derivative_epsilon}
\end{equation}
\begin{widetext}
Using Eqs.~(\ref{eq:grad_area}) and (\ref{eq:grad_perim}) and the chain rule in Eq.~(\ref{eq:derivative_epsilon}), one can show that
\begin{equation}
    \frac{\partial^{2}A_{C}}{\partial R_{k}^{\alpha}\partial \epsilon}=\!\!\!\sum_{\langle m,n\rangle\in C}\! \frac{1}{2} \delta_{C,k} q_{m,n}^{y}\ell_{y} \varepsilon_{\alpha x}(\delta_{m,k}\delta_{n,k+1}+\delta_{m,k-1}\delta_{n,k}),
\end{equation}
and
\begin{equation}
\frac{\partial^{2}P_{C}}{\partial R_{k}^{\alpha}\partial\epsilon} =-\!\!\sum_{\langle m,n\rangle\in C} \delta_{C,k} q_{m,n}^{y}\ell_{y} \left[\delta_{m,k-1}\delta_{n,k} \left(\frac{l_{k-1,k}^{\alpha}l_{k-1,k}^{x}}{\left|\mathbf{l}_{k-1,k}\right|^{3}}-\frac{\delta_{\alpha,x}}{\left|\mathbf{l}_{k-1,k}\right|} \right) 
- \delta_{m,k}\delta_{n,k+1} \left(\frac{l_{k,k+1}^{\alpha}l_{k,k+1}^{x}}{\left|\mathbf{l}_{k,k+1}\right|^{3}}-\frac{\delta_{\alpha,x}}{\left|\mathbf{l}_{k,k+1}\right|}\right)\right],
\end{equation}
where the summation is restricted over all junctions $\langle m,n\rangle$ that belong to cell $C$. Similarly, one can find that
\begin{equation}
\frac{\partial^{2}|\mathbf{l}_{i,j}|}{\partial R_{k}^{\alpha}\partial\epsilon}=
-q_{i,j}^{y}\ell_{y} \left(\delta_{j,k}-\delta_{i,k}\right) \left(\frac{l_{i,j}^{\alpha}l_{i,j}^{x}}{l_{i,j}^{3}}-\frac{\delta_{\alpha,x}}{l_{ij}}\right).
\end{equation}
\end{widetext}


\section{Stress tensor for each cell in the Vertex Model}
\label{appendix:stress}
In this appendix, we provide a detailed derivation of the expression for the stress tensor $\hat{\boldsymbol{\sigma}}_C$ for cell $C$ in the vertex model. Note that the derivation follows the steps introduced in Ref.~\cite{nestor2018relating}, and we further emphasize the difference between internal and external forces to demonstrate that the internal dissipation forces directly produce stresses, while the external dissipation forces produce stresses indirectly via the force balance with internal forces.

In the continuum limit, the mechanical equilibrium can be expressed as $\nabla\cdot\hat{\boldsymbol{\sigma}} + \mathbf{f}^{\text{ext}} = 0$, where 
$\hat{\boldsymbol{\sigma}}(\mathbf{R})$ is the symmetric stress tensor, $\mathbf{f}^{\text{ext}}(\mathbf{R})$ is the external force applied to the system, and $\mathbf{R}$ is a position vector. Note that the vertex model with overdamped dynamics is considered to be in a quasi-mechanical equilibrium.
For a system that is in mechanical equilibrium, we also have $\hat{\boldsymbol{\sigma}}=\nabla\cdot\left(\mathbf{R}\otimes\hat{\boldsymbol{\sigma}}\right) + \mathbf{R}\otimes\mathbf{f}^{\text{ext}}$, where $\otimes$ represents the tensor product. By integrating this relation over an arbitrary area element we obtain
\begin{eqnarray}
\int_{A}\, dA\, \hat{\boldsymbol{\sigma}} &=&\int_{A}\,dA\,\left( \nabla\cdot\left(\mathbf{R}\otimes\hat{\boldsymbol{\sigma}}\right) + \mathbf{R}\otimes\mathbf{f}^{\text{ext}}\right),\nonumber \\
\int_{A}\, dA\, \hat{\boldsymbol{\sigma}} &=&\int_{\partial A} ds\, \mathbf{R}\otimes \left(\hat{\boldsymbol{\sigma}}\cdot\hat{\mathbf{n}}\right)+\int_{A}\, dA\, \mathbf{R}\otimes\mathbf{f}^{\text{ext}},\nonumber \\
\int_{A}\, dA\, \hat{\boldsymbol{\sigma}} &=&\int_{\partial A} ds\, \mathbf{R}\otimes \mathbf{t}+\int_{A}\, dA\, \mathbf{R}\otimes\mathbf{f}^{\text{ext}},\label{eq:stress_general}
\end{eqnarray}
where we used the Stokes theorem to convert the integral over area $A$ into the integral
over the boundary $\partial A$ with the outwards pointing unit normal vector $\hat{\mathbf{n}}$
and the length element $ds$. In Eq.~(\ref{eq:stress_general}) we also introduced boundary traction forces $\mathbf{t}=\hat{\boldsymbol{\sigma}}\cdot\hat{\mathbf{n}}$. Note that if the area element in Eq.~(\ref{eq:stress_general}) is restricted to a subset of the system, then the traction forces $\mathbf{t}$ are resulting from the internal forces between this area element and the rest of the system. Finally, we note that for a system that is mechanical equilibrium the position vectors $\mathbf{R}$ can be measured relative to an arbitrary origin $\mathbf{R}_O$. To demonstrate this, we introduce $\mathbf{R}= \tilde{\mathbf{R}}+\mathbf{R}_O$ and rewrite Eq.~(\ref{eq:stress_general}) as
\begin{eqnarray}
\int_{A}\, dA\, \hat{\boldsymbol{\sigma}} &=&  \int_{\partial A} ds\, \tilde{\mathbf{R}}\otimes \mathbf{t}+\int_{A}\, dA\, \tilde{\mathbf{R}}\otimes\mathbf{f}^{\text{ext}} \nonumber \\
&& + \mathbf{R}_O \otimes \left(\int_{\partial A} ds\,   \mathbf{t}+\int_{A}\, dA\, \mathbf{f}^{\text{ext}}\right), \nonumber \\
\int_{A}\, dA\, \hat{\boldsymbol{\sigma}} &=& \int_{\partial A} ds\, \tilde{\mathbf{R}}\otimes \mathbf{t}+\int_{A}\, dA\, \tilde{\mathbf{R}}\otimes\mathbf{f}^{\text{ext}},
\end{eqnarray}
where we took into account the force balance between boundary tractions and external forces, i.e., $\int_{\partial A} ds\,   \mathbf{t}+\int_{A}\, dA\, \mathbf{f}^{\text{ext}}=0$.

The general formalism discussed above can be applied to a cell $C$ with area $A_C$ in the vertex model to derive the stress tensor $\hat{\boldsymbol{\sigma}}_C$. Before continuing, we introduce the geometric center of the cell $C$ with $N_C$ vertices as $\mathbf{R}_{C}=\frac{1}{N_{C}}\sum_{i\in C}\mathbf{R}_{i}$, where the summation is over all vertices $i$ that belong to the cell $C$. We also define the position of vertices $i$ relative to the cell center $C$ as $\tilde{\mathbf{R}}_{i}=\mathbf{R}_{i}-\mathbf{R}_{C}$. The stress tensor $\hat{\boldsymbol{\sigma}}_C$ for cell $C$ can then be expressed as 
\begin{eqnarray}
    \hat{\boldsymbol{\sigma}}_C &=& \frac{1}{A_C} \int_{A}\, dA\, \hat{\boldsymbol{\sigma}}, \nonumber \\
    \hat{\boldsymbol{\sigma}}_C &=& \frac{1}{A_C}\int_{\partial A} ds\, \tilde{\mathbf{R}}\otimes \mathbf{t}+\frac{1}{A_C}\int_{A}\, dA\, \tilde{\mathbf{R}}\otimes\mathbf{f}^{\text{ext}}.\label{eq:stress_sigmaC_continuum}
\end{eqnarray}
In the vertex model, the degrees of freedom are vertices. Thus the integrals in Eq.~(\ref{eq:stress_sigmaC_continuum}) can be rewritten as
\begin{eqnarray}
\hat{\boldsymbol{\sigma}}_C &=& \frac{1}{A_C} \sum_{i \in C} \tilde{\mathbf{R}}_i \otimes \mathbf{T}_{i\rightarrow C}+\frac{1}{A_C} \sum_{i \in C} \tilde{\mathbf{R}}_i \otimes\mathbf{F}^{\text{ext}}_{i\rightarrow C}. \label{eq:stress_tensor_tractions_external_forces}
\end{eqnarray}
Here, $\mathbf{T}_{i\rightarrow C}$ and $\mathbf{F}^{\text{ext}}_{i\rightarrow C}$ are traction forces and external forces on the cell $C$ due to the vertex $i$, respectively.

\begin{figure*}[t]
    \centering
    \includegraphics[width=0.75\textwidth]{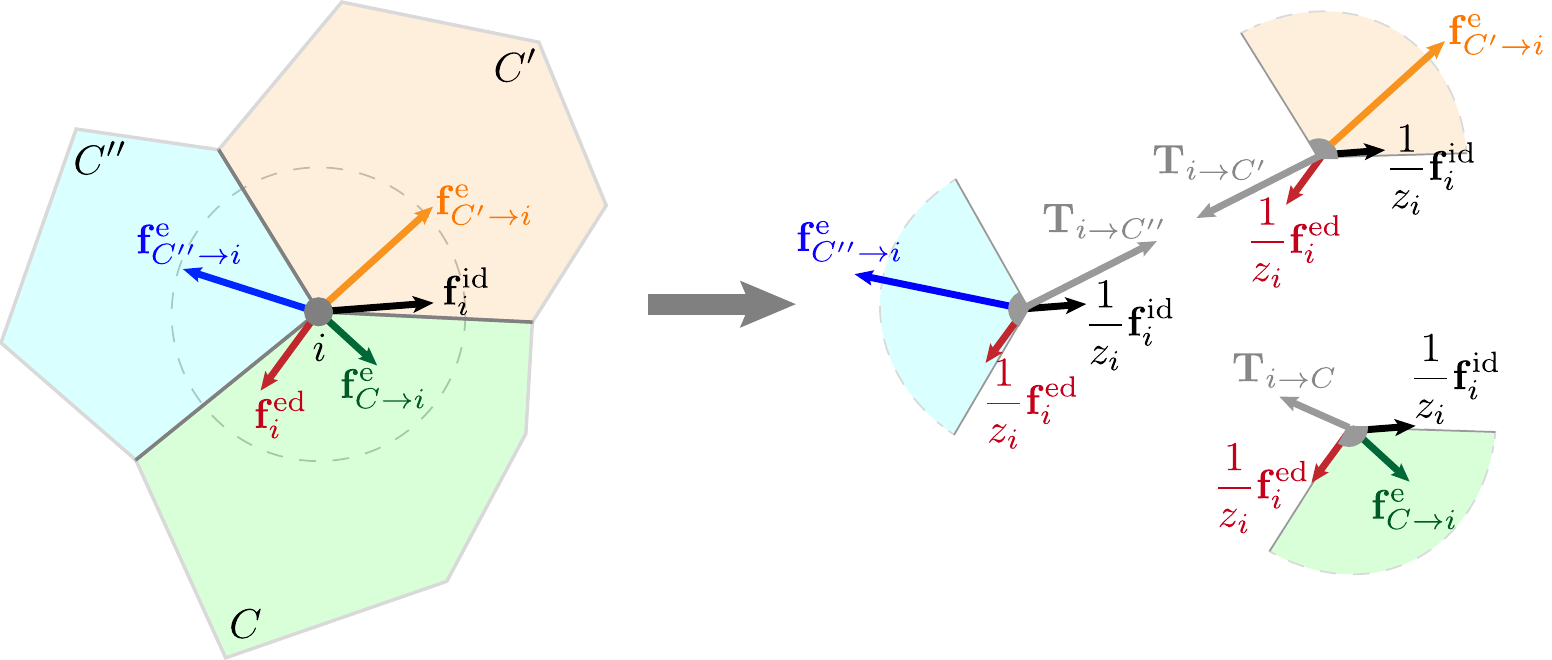}
    \caption{The total force on a vertex is a sum of the mechanical forces from surrounding cells (blue, orange, and green) and internal (black) and external (red) dissipative forces. In order to compute the stress on each cell, it is convenient to make a virtual split of the vertex between all cells sharing it, as shown in the right panel. The force balance is then used for each subvertex to compute the reaction forces (grey) that develop due to the interactions between subvertices.}
    \label{fig:free_body_diagram}
\end{figure*}

Next we discuss how to relate the traction forces $\mathbf{T}_{i\rightarrow C}$ and external forces $\mathbf{F}^{\text{ext}}_{i\rightarrow C}$ to the elastic ($\mathbf{f}_i^\text{e}$) and dissipative  ($\mathbf{f}_i^\text{id}$, $\mathbf{f}_i^\text{ed}$) forces acting on the vertex $i$. First, we note that each vertex $i$ is in a force balance, i.e.,
\begin{eqnarray}
    \mathbf{0} &=& \mathbf{f}_i^\text{e} + \mathbf{f}_i^\text{id} + \mathbf{f}_i^\text{ed}, \nonumber\\
     \mathbf{0} &=& \sum_{C \in \mathcal{N}_i} \mathbf{f}^{\mathrm{e}}_{C\rightarrow i} + \mathbf{f}_i^\text{id} + \mathbf{f}_i^\text{ed},
\end{eqnarray}
where $\mathbf{f}^{\mathrm{e}}_{C\rightarrow i}$ is the elastic force contribution on the vertex $i$ due to cell $C$ and the set $\mathcal{N}_i$ contains $z_i$ cells that share vertex $i$. Then, we draw a free body diagram, where we split each vertex $i$ to $z_i$ subvertices (see Fig.~\ref{fig:free_body_diagram}).
Each subvertex belongs to one of the $z_i$ cells that share the vertex $i$. The subvertex that belongs to the cell $C$ experiences the elastic force $\mathbf{f}^{\mathrm{e}}_{C\rightarrow i}$. For simplicity, we assume that internal and external dissipative forces are distributed equally among the $z_i$ subvertices. Thus, each subvertex experiences the internal dissipative force $\mathbf{f}_i^\text{id}/z_i$ and the external dissipative force $\mathbf{f}_i^\text{ed}/z_i$. Internal forces between subvertices may also develop. The traction force $\mathbf{T}_{i\rightarrow C}$ thus represents the resultant force between the subvertex that belongs to the cell $C$ and all other subvertices. Each subvertex is in force balance and thus we can extract the traction force as
\begin{equation}
    \mathbf{T}_{i\rightarrow C} = - \mathbf{f}^{\mathrm{e}}_{C\rightarrow i} - \frac{1}{z_i}\mathbf{f}_i^\text{id} - \frac{1}{z_i}\mathbf{f}_i^\text{ed}.\label{eq:tractions}
\end{equation}
It is easy to check that the sum of all internal forces between subvertices, i.e., $\sum_{C \in \mathcal{N}_i} \mathbf{T}_{i\rightarrow C} = \mathbf{0}$, vanishes due to Newton's third law. Finally, we note that the external force $\mathbf{F}^{\text{ext}}_{i\rightarrow C}$ arises due to the external dissipation between the subvertex that belongs to the cell $C$ and the substrate, i.e., 
\begin{equation}
\mathbf{F}^{\text{ext}}_{i\rightarrow C} = \frac{1}{z_i}\mathbf{f}_i^\text{ed}.\label{eq:external_force}
\end{equation}

Using the expressions for the traction force $\mathbf{T}_{i\rightarrow C}$ in Eq.~(\ref{eq:tractions}) and the external force $\mathbf{F}^{\text{ext}}_{i\rightarrow C}$ in Eq.~(\ref{eq:external_force}), we can express the stress tensor $\hat{\boldsymbol{\sigma}}_C$ for the cell $C$ in Eq.~(\ref{eq:stress_tensor_tractions_external_forces}) as
\begin{eqnarray}
    \hat{\boldsymbol{\sigma}}_C &=& \frac{1}{A_C} \sum_{i \in C} \tilde{\mathbf{R}}_i \otimes \left ( \mathbf{T}_{i\rightarrow C} + \mathbf{F}^{\text{ext}}_{i\rightarrow C} \right), \nonumber \\ 
  \hat{\boldsymbol{\sigma}}_C &=& \frac{1}{A_C} \sum_{i \in C} \tilde{\mathbf{R}}_i \otimes \left ( - \mathbf{f}^{\mathrm{e}}_{C\rightarrow i} - \frac{1}{z_i}\mathbf{f}_i^\text{id} - \frac{1}{z_i}\mathbf{f}_i^\text{ed}+\frac{1}{z_i}\mathbf{f}_i^\text{ed} \right), \nonumber \\
  \hat{\boldsymbol{\sigma}}_C &=& - \frac{1}{A_C} \sum_{i \in C} \tilde{\mathbf{R}}_i \otimes \mathbf{f}^{\mathrm{e}}_{C\rightarrow i} - \frac{1}{A_C} \sum_{i \in C} \tilde{\mathbf{R}}_i \otimes \frac{1}{z_i}\mathbf{f}_i^\text{id}, \nonumber \\
  \hat{\boldsymbol{\sigma}}_C &\equiv&  \hat{\boldsymbol{\sigma}}_C^\text{e} +  \hat{\boldsymbol{\sigma}}_C^\text{id}.
\end{eqnarray}
In the above equation, the first time corresponds to the stress tensor $\hat{\boldsymbol{\sigma}}_C^\text{e}$ for cell $C$ due to elastic forces, and the second term corresponds to the stress tensor $\hat{\boldsymbol{\sigma}}_C^\text{id}$ for cell $C$ due to internal dissipative forces.

For the vertex model, the elastic force $\mathbf{f}^{\mathrm{e}}_{C\rightarrow i}$ on the vertex $i$ due to cell $C$ is given in Eq.~(\ref{eq:elastic_force_from_cel_to_vertex}) the stress tensor due to elastic forces can be expressed as
\begin{eqnarray}
\hat{\boldsymbol{\sigma}}_C^\text{e} &=& - \frac{1}{A_C} \sum_{i \in C} \tilde{\mathbf{R}}_i \otimes \mathbf{f}^{\mathrm{e}}_{C\rightarrow i},\nonumber\\ 
\hat{\boldsymbol{\sigma}}_C^\text{e} &=&+ \frac{K(A_C - A_0)}{2 A_C} \sum_{i \in C} \tilde{\mathbf{R}}_i \otimes \left[ \left(\mathbf{l}_{i,i+1}-\mathbf{l}_{i,i-1}\right)\times\mathbf{e}_{z}\right] \nonumber \\
&& - \frac{\Gamma\left(P_{C}-P_{0}\right)}{A_C} \sum_{i \in C} \tilde{\mathbf{R}}_i \otimes   \left(\mathbf{\hat{l}}_{i,i-1}+\hat{\mathbf{l}}_{i,i+1}\right), \label{eq:elastic_stress_1} \\
\hat{\boldsymbol{\sigma}}_C^\text{e} &\equiv& \hat{\boldsymbol{\sigma}}_C^\text{e,area} + \hat{\boldsymbol{\sigma}}_C^\text{e,per}.\nonumber
\end{eqnarray}
In the above equation, the first term describes the contribution to the elastic stress tensor $\hat{\boldsymbol{\sigma}}_C^\text{e,area}$ due to the mismatch of the cell area $A_C$ from the target area $A_0$. Via direct calculation of the tensor components, one can show that this stress tensor can be expressed as 
\begin{equation}
\hat{\boldsymbol{\sigma}}_C^\text{e,area}=-\Pi_C\hat{\boldsymbol{I}},
\end{equation}
where $\Pi_C = -\frac{\partial E}{\partial A_C}=-K\left(A_C-A_0\right)$ is the hydrostatic pressure inside the cell $C$ and  $\hat{\boldsymbol{I}}$ is the unit tensor.

The second term in Eq.~(\ref{eq:elastic_stress_1}) describes contribution to the elastic stress tensor $\hat{\boldsymbol{\sigma}}_C^\text{e,per}$ due to the mismatch of the cell perimeter $P_C$ from the target perimeter $P_0$. This term can be further simplified by taking into account that the summation over vertices $i$ is cyclic to write
\begin{eqnarray}
\hat{\boldsymbol{\sigma}}_C^\text{e,per} & = &  - \frac{\Gamma\left(P_{C}-P_{0}\right)}{A_C} \sum_{i \in C} \tilde{\mathbf{R}}_i \otimes   \left(\mathbf{\hat{l}}_{i,i-1}+\hat{\mathbf{l}}_{i,i+1}\right), \nonumber \\
\hat{\boldsymbol{\sigma}}_C^\text{e,per} & = &  - \frac{\Gamma\left(P_{C}-P_{0}\right)}{A_C} \sum_{i \in C}\left( \tilde{\mathbf{R}}_{i+1} \otimes   \mathbf{\hat{l}}_{i+1,i}+ \tilde{\mathbf{R}}_i \otimes  \hat{\mathbf{l}}_{i,i+1}\right), \nonumber \\
\hat{\boldsymbol{\sigma}}_C^\text{e,per} & = &  - \frac{\Gamma\left(P_{C}-P_{0}\right)}{A_C} \sum_{i \in C}\left( -\tilde{\mathbf{R}}_{i+1} + \tilde{\mathbf{R}}_i \right) \otimes  \hat{\mathbf{l}}_{i,i+1}, \nonumber \\
\hat{\boldsymbol{\sigma}}_C^\text{e,per} & = &  + \frac{\Gamma\left(P_{C}-P_{0}\right)}{A_C} \sum_{i \in C}\mathbf{l}_{i,i+1} \otimes  \hat{\mathbf{l}}_{i,i+1}, \nonumber \\
\hat{\boldsymbol{\sigma}}_C^\text{e,per} & = &  + \frac{\Gamma\left(P_{C}-P_{0}\right)}{A_C} \sum_{i \in C}\hat{\mathbf{l}}_{i,i+1} \otimes  \mathbf{l}_{i,i+1}. 
\end{eqnarray}
The final expression above can be rewritten as
\begin{equation}
\hat{\boldsymbol{\sigma}}_C^\text{e,per}  =  \frac{1}{2A_C}\sum_{e\in C}\mathbf{T}_e\otimes\mathbf{l}_e,
\end{equation}
where we introduced the tension $\mathbf{T}_e=\frac{\partial E}{\partial\mathbf{l}_e}=2 \Gamma\left(P_C-P_0\right) \hat{\mathbf{l}}_e$ along the junction $e$, and the summation is over all junctions $e$ that belong to cell $C$. The total stress tensor $\hat{\boldsymbol{\sigma}}^\text{e}_C$ for cell $C$ due to elastic forces can thus be expressed concisely as 
\begin{equation}
\hat{\boldsymbol{\sigma}}^\text{e}_C=-\Pi_C\hat{\boldsymbol{I}}+\frac{1}{2A_C}\sum_{e\in C}\mathbf{T}_e\otimes\mathbf{l}_e.
\end{equation}

Finally, we note that the in the absence of torque on a cell due to internal dissipative forces $\mathbf{f}_i^\text{id}$, the stress tensor $\hat{\boldsymbol{\sigma}}_C^\text{id}$ is symmetric~\cite{nestor2018relating}. This allows us to symmetrize the stress tensor as  
\begin{equation}
\hat{\boldsymbol{\sigma}}_C^\text{id} =  -\frac{1}{2z_i A_{c}}\sum_{i\in C}\left(\tilde{\mathbf{R}}_{i}\otimes\mathbf{f}_{i}^{\text{id}}+\mathbf{f}_{i}^{\text{id}}\otimes\tilde{\mathbf{R}}_{i}\right).
\end{equation}

\bibliography{tissues.bib}
\end{document}